\documentclass[aps,pra,showpacs,preprintnumbers,amsmath,amssymb, 
twocolumn, %tightenlines,
html, reprint
]{revtex4}

\usepackage{color}
\usepackage[bookmarks=true,
bookmarksnumbered=false, % true means bookmarks in
bookmarksopen=false, % true means only level 1
colorlinks=true,
linkcolor=webblue]{hyperref}
\definecolor{webblue}{rgb}{0, 0, 0.5} % less intense blue
\hypersetup{colorlinks=true,urlcolor=webblue,citecolor=webblue}
\usepackage{microtype}

\usepackage[export]{adjustbox}
\usepackage{amsmath,amsfonts,amsthm, amssymb, braket, relsize, exscale, mathtools}
\usepackage{float}
\usepackage{mathrsfs, morefloats}
\usepackage{dsfont}
\usepackage{graphicx}
\usepackage[colorinlistoftodos]{todonotes}
\usepackage{makeidx}
\usepackage{slashed}
\usepackage{lmodern}
\usepackage{feynmp}
\usepackage[caption=false]{subfig}
\usepackage{setspace}
\usepackage{epstopdf}

\begin{document}

\bibliographystyle{apsrev} 

\title {Enhanced creation of high energy particles in colliding laser beams}
%laser standing waves?

\author{M. Yu. Kuchiev} \email[Email: ]{ kmy@phys.unsw.edu.au} \author{J. Ingham}\email[Email: ]{ z3374563@unsw.edu.au}

\affiliation{School of Physics, University of New South Wales, Sydney
  2052, Australia}
    \date{\today}
    \begin{abstract} 
The creation of particles by two colliding strong laser beams is considered. It is found that the electron-positron pairs created in the laser field via the Schwinger mechanism may recollide after one or several oscillations in the field. Their collision can take place at high energy, which the pair gains from the field. As a result, high energy gamma quanta can be created by inelastic scattering or annihilation of the pair. Moreover, heavy particles such as muon pairs may also be created via the annihilation $e^+ + e^-\rightarrow  \mu^+ + \mu^- $. The probability of $e^-e^+$ collision is greatly enhanced due to a strong alignment of the electron and positron momenta with the electric field. The found muon creation rate exponentially exceeds the rate predicted by the direct Schwinger mechanism for muons, while the photon creation rate exponentially exceeds photon emission due to the fermion oscillation.
    \end{abstract}

    \pacs{34.80.Qb, 12.20.Ds, 13.66.De, 23.20.Ra.
          }

    \maketitle

%    \section{introduction}
%    \label{intro}

\section{Introduction}
\label{intro}
This paper examines the probability of particle production by two strong colliding laser beams. The laser frequencies are low, $\hbar \omega\ll m c^2$. However, we find that qualitatively important features of the process deviate from the case of particle creation by a static electromagnetic field originally considered by Sauter, and formulated in the language of quantum field theory by Schwinger 
\cite{Sauter1931,Schwinger1951}. The electron-positron ($e^-e^+$) pairs created in the oscillating laser field may recollide; these collisions violate the adiabatic condition, and consequently strongly enhance the probabilities of a number of phenomena.
The phenomena considered in this paper are the creation of high energy gamma quanta and muon production. The found rate for $\mu^-\mu^+$ production exponentially exceeds
the rate predicted by the Schwinger mechanism, while high energy photon production by low frequency laser fields is a new phenomenon, which has not been discussed previously. 

The $e^-e^+$ pair creation process predicted in \cite{Schwinger1951} for static fields has since become a popular area of study---it lies outside the ambit of perturbation theory and illustrates the nonlinear properties of the QED vacuum. The problem was subsequently extended to electromagnetic fields that vary slowly \cite{Brezin1970,Narozhnyi1973, Marinov1977, Popov1971}, and thereafter, the creation of $e^-e^+$ pairs due to such fields colliding with various targets has been extensively studied in literature \cite{Yakovlev1966, Muller2003, Kuchiev2007a, Ringwald2001, Muller2004, Shearer1973, Liang1998,  Muller2008, Nerush2011, Elkina2011, Alkofer2001, Muller2008a}.  

Despite its theoretical import, pair production due to static, or slowly-varying strong fields has only been indirectly observed \cite{Bamber1999,Narozhnyi2015} as it is strongly suppressed for experimentally achievable electric fields. The critical field strength for particle creation is $\mathcal E_c = m^2c^3/e\hbar=  1.32 \times 10^{16}$ V/cm,
requiring an intensity of 
$\text{I}=c \mathcal E_c^2/(8\pi)= 2.33 \times 10^{29}$ W/cm$^2$, 
much larger than the fields available experimentally.
%m = 0.511 10^6;
%r = 0.5292/137.036 10^-8;
%q = 1.602 10^-19;
%v = 3 10^10;
%\[Alpha] = 1/137.036;
%field = m/r V/cm
%flux = (m  q)/(8 \[Pi] r^3) v/\[Alpha]   W/cm^2
However, in recent years, several large laser facilities including ELI \cite{ELI2011}, XFEL \cite{xfeleu} and HiPER \cite{hiperlaser}, have come increasingly close to creating the critical electric field strength required for observing pair production---HiPER, for instance, is predicted to produce intensities of $10^{26}$ W/cm$^2$. These theoretical and experimental advances mean laser facilities may test theories of strong field QED in the next decade, motivating further study into what they may observe.

Several contemporary efforts are therefore directed at means to improve the observability of pair production.  Recent works investigate manipulation of the laser pulse shape \cite{Kohlfurst2013} and tuning of the laser polarization \cite{Kirk2009}. Many studies have investigated the so-called dynamical Schwinger mechanism \cite{Schutzhold2008}, brought about by colliding two laser beams--- one strong low frequency field (which gives rise to Schwinger pair production), and a weak high frequency field which lowers the barrier for the strong field, Schwinger-type process. Recent work by Di Piazza et al. \cite{DiPiazza2009} claims that by tuning the frequency of the weak field, $e^-e^+$ creation may be observable with presently available technology. 

Another means to enhance pair production provides the so-called antenna mechanism \cite{Kuchiev2007}, 
%which enhances the creation of high energy particles provided a weak $e^-e^+$ creation rate. 
which is a three step process. First, an $e^-e^+$ pair is created through the Schwinger mechanism. Second, the $e^-$ and $e^+$ oscillate in the laser field, and third they recollide. The pair has gained energy by accelerating in the field, and this energy becomes available for the creation of new particles upon collision. In the original study \cite{Kuchiev2007}, $e^-e^+$ pairs created by the collision of a heavy nucleus and laser beam were considered, and it was shown that high energy inelastic $e^-e^+$ collisions are possible. In a recent paper by Meuren et. al. \cite{Meuren2015}, the recollision probability was found in the case of a single photon colliding with a strong optical laser pulse and was demonstrated to contain subleading dependence on the laser frequency \cite{Meuren2015a}. 
%Here, we shall find an analytical formula for the recollision probability that possesses a similar subleading frequency dependence.

\begin{figure}[h]
\vspace*{-2.4cm}
\hspace*{-2cm}
\includegraphics[scale=0.555]{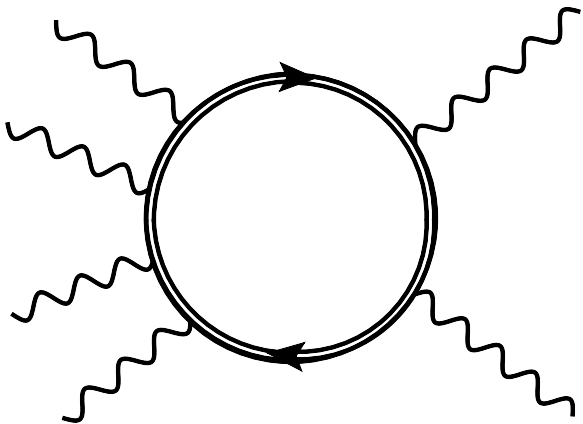}
\caption{Diagram for an $N\gamma \rightarrow e^-+e^+\rightarrow2\gamma $ process, with $N=4$. The double lines represent the laser-dressed electron propagators; wavy lines on the left represent low frequency laser photons, while those on the right represent high frequency photons. In the processes we consider, $N=m/\omega\gg 1$.}
 \label{Feynman2005diag}
\end{figure}

The complete antenna-type process can be written as
\begin{align}
\label{mis}
n\gamma\rightarrow e^- + e^+ \rightarrow X^- + X^+,
\end{align} where $X$ may be any of a number of particles. For example, $X$ can be an electron, in which case the last step in \eqref{mis} describes scattering $e^- + e^+ \rightarrow e^- + e^+ $. During this process the $e^-e^+ $ pair fulfils a transition from one adiabatic state in the laser field to another one. The antenna mechanism then provides an efficient exchange of energy between the pair and the field. Another possibility is that $X$ is a photon, in which case \eqref{mis} describes high energy photon production, see Figure \ref{Feynman2005diag}. Ultimately, we will show that the antenna may accumulate enough energy for $X$ to be essentially any particle that interacts with electrons. 

The first transition in \eqref{mis} is a multiphoton transition by a strong field, and hence exponentially suppressed, in accordance with the Schwinger result. However, the second transition is a non-adiabatic inelastic collision, and therefore exhibits only power-type suppression. This is the salient point: the creation rate is enhanced by replacing exponential suppression of the high energy particle $X$ with power-law suppression.

An analogy with atomic physics is instructive. The analogue of the $e^-e^+$ antenna in atomic physics is the atomic antenna originally proposed in \cite{Kuchiev1987} and described semiclassically by Corkum, and Lewenstein et al. \cite{Corkum1993, Lewenstein1994}. A laser wave ionizes an atom, accelerates the free electron, before colliding the electron with its parent atomic particle---producing high-harmonic radiation, multiple ionization and above threshold ionization \cite{Lewenstein1994, Kuchiev2000, Kuchiev1999}. This is currently an area of active research \cite{Shwartz2014, Smirnova2009}.
\newpage
Schwinger showed in \cite{Schwinger1951} that the probability rate per volume of fermion pair creation $\dot{W}_s$ due to a static electric field of strength $\mathcal E$ is \begin{align}\label{Schwinger1951}
{W}_s = \frac{\mu^4}{4\pi^3}\left(\frac{e\mathcal E}{\mu^2}\right)^2\exp\left( -\frac{\pi \mu^2}{e\mathcal E} \right).
\end{align} 
where $\mu$ is the fermion mass. For large masses, this process is therefore drastically suppressed as compared with electron production. Remarkably, the probability rate due to the antenna mechanism considered in our work is 
\begin{align}\label{introant}
{W}_a = \Upsilon~\frac{m^4}{4\pi^3}\left(\frac{e\mathcal E}{m^2}\right)^2 \exp\left(-\frac{\pi m^2}{e\mathcal E}\right)
\end{align} where $m$ is the electron mass, and 
the factor $\Upsilon$
incorporates power-law dependence on $m/\mu $. Whilst the exponent in \eqref{Schwinger1951} depends on the mass $\mu$ of a heavy particle, the exponent in \eqref{introant} depends on the electron mass $m$, and dependence on $\mu$ only appears preeponentially. In such a way, the probability \eqref{introant} is exponentially larger than \eqref{Schwinger1951} \footnote{
For muon creation, the ratio between \eqref{Schwinger1951} and \eqref{introant} is enormous---roughly $\exp \left( \varphi ({m_\mu}/{m})^2\right)\approx 10^{\, 2\varphi \times 10^5}$, where $\varphi=\pi m^2/({e\mathcal E})\lesssim 1$.}. We also predict emission of photons with high frequency $\Omega \gg m $, whose creation rate is also described by \eqref{introant}, for which the exponent includes the electron mass and does not incorporate $\Omega$. We present formulae for the probability rates and spectra of the recollision processes, for laser frequencies $\omega \ll m$ and large adiabatic parameter 
$\xi\gg 1$. Interaction between the $e^-e^+$ pair and the laser field is included to all orders via the dressed fermion wave functions.

Interestingly, the preexponential factor $\Upsilon$, which determines the rate of $e^-e^+$ annihilation, proves to be much larger than classical estimates anticipate due to the fact that the velocities of the created $e^-$ and $e^+$ are correlated along the direction of the laser electric field, enhancing the yield of quantum process $W_\text{quant}$ compared to the classical estimate $ W_\text{class}$ by a factor
\begin{align}
\frac{ W_\text{quant}}{W_\text{class} }\propto \frac{c^3}{\hbar}\frac{\xi^2}{ \ln^2(\pi\xi)}
\frac{m^2}{e\mathcal E} \gg 1.
\label{quantum-class}
\end{align}
Here $\xi=e\mathcal E/(m\omega)$ is the adiabatic parameter, 
which is presumed large, $\xi \gg 1$, and 
the strength of the considered electric field is also small, ${e\mathcal E} \lesssim m^2$.

Section II of this paper describes parameters of the laser field we consider. Section III presents the probability of $e^-e^+$ pair creation. Section IV derives the wave function of the $e^-e^+$ pair created in a laser field, which allows us to derive the recollision probability in Section VI. Section V discusses the influence of photon radiation during the electron propagation. Section VII presents an analytical estimate for the creation rates in Section VI, which is then compared with numerical calculations in Section VIII. Section VIII presents the spectra and total creation rate for photons produced via recollisions. Section IX discusses the possibility of muon production, and presents an approximation for the relevant creation rates. Natural units $\hbar=c=1$ are used.	
\section{Assumptions}
\label{assumptions}
The electric fields considered below are presumed to be weak, 
\begin{align}\label{weakfield}
F= e {\mathcal E} \lesssim e {\mathcal E}_c = m^2,
\end{align} 
a condition satisfied by the presently available laser fields \cite{Narozhnyi2015, Dunne2009, Heinzl2011}. 
Furthermore, we presume that the laser frequency $\omega$ is small, $\omega \ll m$. 
The latter presumption implies that the creation process requires the absorption of a large number of photons. 

The behavior of electrons in the laser field is governed by an important dimensionless quantity,
the afore-mentioned adiabaticity parameter $\xi$ 
\footnote{Sometimes the process is described using the parameter $\gamma=\xi^{-1}$.}
\begin{align}
\xi = \frac{\mathcal E}{\mathcal E_c}\frac{m}{\omega}=\frac{F}{m\omega}\gg 1,
\label{xi}
\end{align}
We consider $\omega$ and $\mathcal E$ within the adiabatic limit, in which $\xi$ is large---as indicated on the right-hand side of (\ref{xi}) (and was mentioned in Eq. (\ref{quantum-class})).
%\begin{align}\label{tunneling}
%\gamma \ll 1.
%\end{align}
%\begin{align}
%\zeta=\frac{1}{\gamma}=\frac{e\mathcal{E}}{m\omega}.
%\end{align}
To illustrate how large $\xi$ may be, take the following numerical values as an example: $\omega=1 \ \text{eV}$ and $e\mathcal E=0.1 m^2$ (these are feasible for lasers currently being developed \cite{Dunne2009}); then
$\xi \approx 5\cdot 10^{4}$.
%\begin{align}
%\label{example}
%\gamma \approx 2\cdot 10^{-5}.
%\end{align} 
%These parameters are feasible for lasers currently being developed, see \cite{Dunne2009}. 
%Same estimate for $\gamma$ is valid for $\omega=10$ eV  and $e\mathcal E=m^2$, which may become
%available in the near future.

%\begin{align}
%\label{example2}
%\gamma\sim 10^{-4}.
%\end{align}

The electric field is supplied as follows.
Two laser pulses collide in the vicinity of some point we choose as the origin. The waves propagate in opposite directions, colliding head-on with equal frequency $\omega$, identical linear polarization and intensity. 
We choose the reference frame in which the direction of propagation is
along the $x$ axis, the electric field points along the $z$ direction and the magnetic field along the $y$ direction.
In addition, we presume that the phases of the beams are tuned in such a way that the electric field reaches its maximum 
at the origin, while the magnetic field is cancelled there.

Altogether, these assumptions mean that we
consider a linearly polarized standing wave that produces the following electric and magnetic fields \footnote{Laser beams must be tightly focused to acquire such strengths, however, the laser field can be modeled as homogeneous within the focal spot of the laser, which is large with respect to the Compton wavelength \cite{Narozhnyi2015}.}
\begin{eqnarray}
\nonumber
\textbf{E}_1(t)+\textbf{E}_2(t) &=& \textbf{E}\cos\omega t \cos kx ~.
%\approx \textbf{E}\cos\omega t,
\label{fields}
\\
\label{mfields}
\textbf{B}_1(t)+\textbf{B}_2(t)&=& -\textbf{B}\sin\omega t \sin kx  \approx 0.
\end{eqnarray} 
%%%%%%
%\begin{figure}[H]
%\centering
%\includegraphics[scale=0.4]{laser_fields.png}
%\caption{Colliding laser beams}
% \label{collision}
%\end{figure}
It is shown in Section IV that after the $e^-e^+$ pair is created,
it propagates along a trajectory close
to the planes $\omega x=m \pi$ , $m=0,\pm 1\,\dots$, where the electric field is large.
As a result, we may neglect  
the magnetic field, as per the right hand 
side of Eq. \eqref{mfields}
\footnote{To make certain that the magnetic field is unimportant, we performed analytical calculations verifying that the pair creation rate due to the electric field remains larger than that due to the magnetic field, even when the strength of the magnetic field is comparable with the electric field. Physically, this can be explained by the fact that there is no tunneling \textit{regim\'e} for the magnetic field.}.
%\begin{figure}[htp]
%\centering
%\includegraphics[scale=0.40]{laser_fields.png}
%\caption{Electric and magnetic fields of the colliding lasers in Eq. \eqref{fields} and \eqref{mfields}.}
% \label{collision}
%\end{figure}
\section{Electron-positron pair creation}
\label{distributions}
We shall summarize several important features of the $e^-e^+$ creation process.
Consider the electric field \eqref{fields}. The slow variation of this field allows us
to describe the probability of pair creation by a simple modification of Eq. (\ref{Schwinger1951}),
substituting there the slow varying electric field instead of the static one.
The exponential dependence in Eq. (\ref{Schwinger1951})
on the electric field implies that the dominant contribution to the creation rate comes from those regions of space-time 
where the field reaches its maximum. This implies that the $e^-e^+$ pairs are created mainly 
in the vicinity of planes $x=n\pi/\omega$.
For convenience we will consider the contribution of the maximum located at $\omega x=\omega t=0$.

To find the averaged probability rate of $e^-e^+$ creation,
we average the Schwinger result (\ref{Schwinger1951}) over the coordinates and time in the vicinity of $\omega x=\omega t=0$,
\begin{align}
\nonumber
\langle W\rangle &=\frac1{\pi^2}\iint_{-\pi/2}^{\pi/2}  W_s(F \cos k x \cos \omega t)  \,{ d(k x) } \,{d(\omega t)}
\\
& \approx \frac{m^4}{2\pi^5}\left(\frac{F}{m}\right)^3\,\exp\Big(\!-\frac{\pi  m^2 }{F}\Big)~.
\label{average-w}
\end{align}
This approximation is arrived at by noting that a small deviation of $x$ and $t$ from zero
results in a rapid decrease of the integrand, and so we may expand the field over $x$ and $t$ about $\omega x \approx \omega t \approx 0$,
\begin{equation}
\!\!W_s(F \cos k x \cos \omega t)\! \approx\! {W_s}(F) \exp \! \big(\! -\pi m^2 \omega^2 \frac {x^2+t^2 }{2F}\big),
\label{expandSCH}
\end{equation}
and extend the limits  of integration to infinity. This technique is commonly employed in the theory of strong-field atomic ionization \cite{Delone2000}.
Altogether, this brings us to the final expression Eq. (\ref{average-w}).

The momentum distribution of created $e^-e^+$ pairs may be found
in the amplitude that describes the
creation of the pair. 
The calculation of this amplitude, which relies on the 
adiabatic, Keldysh-type approach \cite{Keldysh1965,Popov1973} is outlined in Appendix \ref{Keldysh method}. The adiabatic nature of the process allows one to express
the amplitude $A(\mathbf{p})$ of pair creation via the 
tunneling action $S_\text{tun}(\mathbf p)$
\begin{equation}
A(\mathbf p)=A_0 \exp(i S_\text{tun}({\mathbf p}))~,
\label{AexpS}
\end{equation}
where $A_0 $ is a preexponential, $\mathbf p $-independent factor, 
while according to Eqs. (\ref{AfromS}) and (\ref{BfromS}), 
the tunneling action reads 
\begin{align}
S_\text{tun}({\mathbf p})&=\frac{i\pi}{2F}\,\Big(m^2+{p_\perp^2}+\frac{ p_\parallel^2}{2\xi^2}\Big)-\frac{1}{2F}\int_0^{p_\parallel}\varepsilon_p(0) \ dp
\label{S(p)}
\end{align}
where $\varepsilon_p(t)$ is the time dependent energy of the electron or positron oscillating in the electric field, which is defined in detail below, see Eq. \eqref{elenergy}. 
In Eq. (\ref{S(p)}) it is presumed that $|p_\parallel|\gg m$, c.f. Eq. (\ref{BfromS}).
Using this result, the probability of the pair creation may be written as follows
\begin{align}
\label{momentumdistribution}
\frac{d^2\langle W \rangle}{dp_\perp dp_\parallel}=&|A(\mathbf{p})|^2 = 
\frac{\sqrt \pi}{2\sqrt 2 \,\xi F^{3/2}}\langle W \rangle 
\\
\nonumber
&\times\exp \Big(\!-\frac{\pi }{{F}}\big( p_\perp^2 + \frac{1}{2\xi^2} p_\parallel^2 \big)\Big),
\end{align}
The exponential dependence on  $\mathbf p$ follows from Eqs. (\ref{AexpS}) and (\ref{S(p)}), while
the $\mathbf p$-independent factor in the first line is determined by requiring the average creation rate comply with Eq. (\ref{average-w}) such that
\begin{equation}
\int\frac{d^2\langle W\rangle}{dp_\perp dp_\parallel}\,d^2p=\langle W\rangle.
\label{intw=w}
\end{equation}
Equation (\ref{momentumdistribution}) can also be verified using the results of Marinov and Popov \cite{Marinov1977}, 
who considered the problem for arbitrary  $\gamma=\xi^{-1}	$. 
In the limit $ \gamma\ll 1 $, their result
reproduces the exponent of \eqref{momentumdistribution}. 

The quantity $\mathbf p$ which appears in Eqs. (\ref{AexpS}), (\ref{S(p)}) and (\ref{momentumdistribution}),
is the momentum of the electron at the moment of creation, $t=0$.
The positron momentum equals $-\mathbf  p$.
The components of the momentum may take any values, positive or negative, though
according to Eq. (\ref{momentumdistribution}) very large momenta are suppressed.
Examination of the exponent in (\ref{momentumdistribution}) shows 
that the component of the electron (positron) momentum collinear with the electric field
is much larger than the momentum transverse to the field,
\begin{equation}
p_\parallel \sim \xi \sqrt F \gg p_\perp \sim  \sqrt F~.
\label{paraperp}
\end{equation}
Additionally, the electron and positron polarizations may be described by
spiral states i.e. eigenstates of the spin projection along the fermion momentum. In this representation, the electron and positron must have opposite spin projections. 
This means the projection of the pair's total spin, along the momentum of one fermion, is zero. 
To verify this, recall that the adiabatic pair creation process in the homogeneous electric field can be understood, similarly to
the static Schwinger process, using an effective
Lagrangian which describes the influence of the field on the vacuum
see e.g. Ref. \cite{Berestetskii1982}. This Lagrangian possesses only trivial quantum numbers. 
Hence, the created pair can have only zero projection of total spin (see also \cite{DiPiazza2012}). By the same token,
the total momentum of the pair also remains zero, as we have already noted. 

The spin polarization of the $e^-e^+$ pair leads to a slight, power-type suppression of the $e^-+e^+\rightarrow X^++X^-$  process \cite{Berestetskii1982} that must be taken into account when calculating the total creation rate of the antenna process in Section \ref{totalprob}.

\section{Pair propagation amplitude}
\label{Quantum description}

In this section we calculate the wave function
$\Psi(\mathbf r_1,\mathbf r_2,t)$ of the
$e^-e^+$ pair which is firstly created, and then
propagates in the electric field, see Section \ref{wavefunction of the pair}.
Our derivation relies on the adiabatic and semiclassical approximations, which 
remain valid since the
frequency of the field is small, $\omega\ll m$ .

The desired wave function describes electron oscillations in the laser field, the magnitudes of which are of the order of $\lambda=2\pi/\omega$. Our objective is to describe the 
collision of the two fermions with high energy $\varepsilon \ge 2m\gg \omega$, 
which results in the creation of
new particles. The collision takes place
in the region of small fermion separations, 
$\delta r_{12}\sim 1/\varepsilon \ll \lambda=2\pi/\omega$, over which
the function $\Psi(\mathbf r_1,\mathbf r_2,t)=\Psi(\mathbf r_{12},t)$
does not show substantial variation.
Accordingly, the $e^-e^+$ collision can be described
by the wave function $\Psi(0,t)$, i.e. the wave function at $r_{12}=0$.
This is analogous to the known problem of the positronium annihilation, 
which is resolved through calculation of $\psi_{pos}(0)$, see e.g.
Ref. \cite{Berestetskii1982}.

The $e^-e^+$ pair undergo a collision, which produces new particles. Generally, $\Phi(0,t)$ is a sophisticated function of $t$,
which distinguishes the present problem from positronium annihilation
where the fermion wavefunction exhibits simple oscillations $\propto \exp(-iE_{Ps}t)$, where $E_{Ps}$ is the total positronium energy. 
However, the complicated $t$-dependence of the wave function reveals itself only at large intervals of time $\delta t\propto 1/\omega$, 
while the annihilation is a rapid process. Hence, the only contribution relevant to the annihilation
originates from the part of wave function that oscillates, at the moment of collision, with the frequency matching the frequency $2\varepsilon$ of the particles (say, photons) in the final state. To extract this term, we need to take the Fourier transform,
\begin{equation}
\Psi(0,t)=\int e^{-i\,2\varepsilon t}\,\tilde\Psi(0,2\varepsilon)\,\frac{d\varepsilon}{\pi} 
\label{int dE}
\end{equation}
where $\tilde\Psi(0,2\varepsilon)$ is the Fourier component of $\Psi(0,t)$.
In order to describe the creation of the final
pair with energy $2\Omega$ we consider the Fourier
component $\tilde\Psi(0,2\Omega)$, that is, the coefficient in front of the term exhibiting the necessary oscillations $\propto e^{-i\,2\Omega t}$.

We conclude that $\tilde\Psi(0,2\Omega)$
is the primary object of interest in describing
the inelastic collision of
$e^-e^+$ pairs, with energy $2\Omega$.
We find $\Psi(0,t)$ in Section \ref{Saddle point approximation}, and
the Fourier component $\tilde\Psi(0,2\Omega)$ in Section \ref{Fourier component of wavefunction}.

\subsection{Wave function of the pair}
\label{wavefunction of the pair}
Let us calculate the wave function which describes the created $e^-e^+$ pair. 
The exact Volkov wave function is not applicable in our case, because the laser standing wave 
considered \eqref{fields} is not a plane wave. However, we can derive a reliable approximation for the wave function using the fact that the electron (positron) momentum and energy greatly exceed the frequency of the field, $|p(t)|,\,\varepsilon(t) \gg \omega$.
Consequently, the motion of the pair in the smooth field can be described using the 
semiclassical approximation, which relies on the classical trajectories of the fermions. From
the previous discussion we know that most of these trajectories are located in the
region where the magnetic field is greatly suppressed, while the electric field is almost homogeneous. 

The approximation corresponding to our problem, of fermion propagation in a homogeneous adiabatic electric field, is detailed in Appendix \ref{Adiabatic approximation}. From the Appendix, the single-particle wave functions are
\begin{align}
\psi^{ (\pm) }_{\mathbf{p}}(\mathbf r,t)=
\chi^{ (\pm) }_{\mathbf{p}(t)} 
\exp\Big(i \big(\,\mathbf{p}(t)\cdot \mathbf r-\int^t\varepsilon(t')dt'\,\big)\Big)~,
\label{psie}
\end{align}
(the superscripts, $(-)$ and $(+)$, refer to the electron and positron states respectively).
The electron (positron) energy
\begin{align}
\varepsilon(t)=(m^2+p^2(t))^{1/2}~,
\label{elenergy}
\end{align}
is related to the electron momentum given by
\begin{align}
\mathbf{p}(t)=\mathbf{p}-e\, \mathbf{A} (t)= \mathbf{p}+\frac{\mathbf F}\omega\sin \omega t,
\label{p(t)-eA}
\end{align}
where $\mathbf A(t)$ is the vector potential $\mathbf F\cos\omega t=-e \dot{\mathbf A}$. In \eqref{psie}, we took into account the fact that
the classical momentum of the positron is opposite to that of the electron when they are created as a pair.
%The physical meaning of Eq. (\ref{psie}) is clear. At each moment of time this wave function describes a plane wave, which parameters smoothly vary with slow-varying electric field. As is usual in the adiabatic approximation, the only point where the additional accuracy is needed is the large phase factor, where the integral of the energy appears.

The factor $\chi^{ (\pm) }_{\mathbf{p}(t)} $ in \eqref{psie} is the Dirac spinor, defined in Appendix \ref{Adiabatic approximation}. At any given moment of time $t$ this spinor may be considered as describing a plane wave with
momentum $\boldsymbol p (t)$ and energy $\varepsilon(t)$. Its distinction from the 
conventional plane wave in vacuo is that its momentum and energy are slowly varying functions of time. 
Overall, the representation of the wave function in (\ref{psie}) complies with conventional machinery of the adiabatic description.  As is usual, 
%the only aspect of the wave function which requires additional accuracy is 
the large phase factor $\int^t\varepsilon(t')dt'$ needs to be written accurately, 
incorporating integration over different moments of time.

To construct the wave function of the $e^-e^+$ pair $\Psi({\mathbf r}_1,{\mathbf r}_2,t)$, which is created and then propagates in the laser field,
it is convenient to interpret this wave function as the amplitude of 
transition from the vacuum state $|\text{vac}\rangle$ into the state 
$|\mathbf r_1,\mathbf r_2,t\rangle$,
in which at the moment of time $t$, the electron and positron have coordinates $\mathbf r_1$ and $\mathbf r_2$. In other words,
\begin{equation}
\Psi({\mathbf r}_1,{\mathbf r}_2,t)=\langle \mathbf r_1,\mathbf r_2,t|\text{vac}\rangle .
\label{psi12t}
\end{equation}
We expand this amplitude in the complete set of states 
$|\mathbf p,-\mathbf p,0\rangle$, i.e. those for which the electron and positron have the momenta 
$\mathbf p$ and $-\mathbf p$ respectively at $t=0$,
\begin{equation}
\langle{ \mathbf r_1,\mathbf r_2,t|\text{vac}}\rangle=\int\langle \mathbf r_1,\mathbf r_2,t|\mathbf p,-\mathbf p,0\rangle
\langle \mathbf p,-\mathbf p,0|\text{vac}\rangle \,\frac{d^3p}{(2\pi)^3}.
\label{psi12tpp0}
\end{equation}
For simplicity of discussion we suppress the reference to the spin projections for now.
Since the electron and positron propagate independently after their creation, 
the amplitude of their joint propagation is a product of two independent amplitudes
\begin{equation}
\langle \mathbf r_1,\mathbf r_2,t|\mathbf p,-\mathbf p,0\rangle=
\langle \mathbf r_1,t|\mathbf p,0\rangle\langle \mathbf r_2,t|-\mathbf p,0\rangle.
\label{multiply}
\end{equation}
Note that $\langle \mathbf p,-\mathbf p,0|\text{vac}\rangle$ is identical to the amplitude of the
pair creation, $\langle \mathbf p,-\mathbf p,0|\text{vac}\rangle=A(\mathbf p)$ from Eq. \eqref{AexpS}, 
while the amplitudes of fermion propagation coincide with fermion single-particle wave functions
\begin{equation}
\langle \mathbf r,t|\pm {\mathbf p},0\rangle=\psi^{ (\mp) }_{\pm\mathbf{p}}(\mathbf r,t).
\label{<>=psi}
\end{equation}
Combining Eqs. \eqref{psi12t}$-$\eqref{<>=psi},  
we present the wave function of the created $e^-e^+$ pair as the following wave packet
\begin{equation}
\Psi({\mathbf r}_1,{\mathbf r}_2,t)\!=\! \int 
A(\mathbf p)\psi^{ (-) }_{\mathbf{p}}(\mathbf r_1,t)\psi^{ (+) }_{-\mathbf{p}}(\mathbf r_2,t)\frac{d^3p}{(2\pi)^3}.
\label{Apsipsi}
\end{equation}
This result may be verified using simple properties of quantum amplitudes,
see e.g. \cite{Feynman2005}.
The amplitudes of independent events
are multiplied, as per the three factors in the integrand of \eqref{Apsipsi}, while
the amplitudes of interfering events need to be added,
which is taken care of by the integration over momenta.

It is convenient to rewrite the wave function of the pair using Eqs. \eqref{AexpS}, \eqref{S(p)} and \eqref{psie} as follows,
\begin{align}
&\Psi({\mathbf r}_{12},t)= A_0 \int  \xi_{\boldsymbol{p}}\, \zeta \,\exp\big(i S(\mathbf p, \mathbf r_{12},t)\big)\,\frac{d^3p}{(2\pi)^3}~,
\label{Phi exp iS}
\end{align}
where
\begin{align}
&\zeta(\boldsymbol{p}) = \exp\Big(-\frac \pi {2F}
(p_\perp^2+\frac1{2\xi^2} p_\parallel^2 )\Big),
\label{chi}
\\
&S(\mathbf p, \mathbf r_{12},t)=\mathbf{p}(t)\cdot \mathbf r_{12}-2\int^t\varepsilon(t')dt'-\frac{1}{F}\int_0^{p_\parallel}\varepsilon_p(0) \ dp~.
\label{Spropagate}
\end{align}
The function 
\begin{align}
\xi_\mathbf p =\hat P (\chi^{ (-) }_{\mathbf{p}(t)}\chi^{ (+) }_{-\mathbf{p}(t)})
\label{PhiNew}
\end{align}
is the spin part of the wave function; $\hat P$ is the projection operator that separates only those spin states of the pair, with zero spin projection on the direction of the pair propagation.

The constant $A_0$ in \eqref{Phi exp iS} is a normalization factor, into which we absorbed the factor of 
$\exp(-\pi m^2/2F)$ present in $A(\mathbf p)$.
We choose
\begin{equation}
A_0^{-2}=\int\zeta^2(\boldsymbol{p})~\frac{d^3p}{(2\pi)^3}=\frac{\xi F^{3/2}}{4\sqrt2\,\pi^3}
\label{norma}
\end{equation}
to ensure the conventional normalization
\begin{equation}
\langle \Psi|\Psi\rangle=1~.
\label{Psi^2=1}
\end{equation}
The spinor part of this identity,
\begin{equation}
\xi_{\boldsymbol{p}}^*\,\xi_{\boldsymbol{p}} =1,
\label{PhiPhi}
\end{equation}
follows from the chosen normalization of spinors in \eqref{uu}. The distance $\mathbf r_{12}=\mathbf r_{1}-\mathbf r_{2}$ is the fermion separation.

The first and second terms in $S$, due to the wave functions \eqref{psie},
constitute the classical action which describes propagation of the pair in the external electric field.  
The third term in $S$ originates from the phase factor in $A(\mathbf p)$, see
\eqref{AexpS} and \eqref{S(p)}. We will discuss the physical significance of this term in the next section.

Summarizing, the wave function of the created $e^-e^+$ pair is found in a simple, appealing form
\eqref{Phi exp iS}.

\subsection{Saddle point approximation}
\label{Saddle point approximation}

We aim to evaluate the wave function of the 
electron-positron pair \eqref{Phi exp iS}
in a convenient form
using the saddle-point
approximation, the method of stationary phase, more specifically. The ratio between the fermion energies and
the frequency of the laser plays the role of large parameter, which justifies this approach.

The action $S$ in \eqref{Spropagate} 
is a rapidly varying function of momentum.
By contrast, $\xi_{\boldsymbol{p}}$ and $\zeta$ from \eqref{PhiNew} and \eqref{chi} are slowly varying functions.
The latter statement is obvious for the preexponential factors
$\chi^{ (-) }_{\mathbf{p}(t)} $ and $ \chi^{ (+) }_{\mathbf{p}(t)} $, which appear in the spin part. 
At a first glance, the function $\zeta$ \eqref{chi}
incorporates a potentially dangerous exponential dependence on $\mathbf p$, which may induce 
rapid variations. However, this function in fact remains slow-varying, as shown at the end of this section.

Evaluating the integral at the saddle point, we find
\begin{equation}
\Psi(\mathbf r_{12},t)\approx 
A_0\, \xi_{\boldsymbol{p}}\,\zeta(\boldsymbol{p})\int \,\exp\big(i S \big)\,\frac{d^3p}{(2\pi)^3}~,
\label{wf-saddle-point}
\end{equation}
where $\xi_{\boldsymbol{p}}$ and $\zeta$ are taken at the saddle point, the location of which is given by the condition
\begin{equation}
\nabla_{\mathbf p}S(\mathbf p, \mathbf r_{12},t)=0
\label{saddle point}
\end{equation}
which leads to
\begin{equation}
(r_{12})_i=2\int_0^t v_i(t') \,dt'+\,2\delta_{i3}\,{\varepsilon(0)}/{F}.
\label{dSdp}
\end{equation}
Here $i=1,2,3$ index spatial coordinates, $p_3=p_\parallel$, while
$v_i(t)={\partial \varepsilon (t)}/{\partial p_i}$ denotes the electron velocity and $\varepsilon(0)$ is the initial electron energy.
In Eq. \eqref{saddle point} we recognize the classical Hamilton-Jacobi equation, so that Eq. \eqref{dSdp} represents the classical trajectory of the pair, expressing the 
electron-positron separation $\mathbf r_{12}(t)$ through the integral of the relative velocity $\mathbf v_{12}(t)=2\mathbf v(t)$. This interpretation shows that the last term in Eq. \eqref{dSdp} equals the separation between the fermions at the moment of creation, $(r_{12})_i=2\delta_{i3} \varepsilon(0)/{F}$.

This term in \eqref{dSdp}
originates from the tunneling amplitude, so we conclude that the initial separation
$2{\varepsilon(0)}/{F}$ between the fermions is produced by the tunneling process, exactly analogous to the tunneling distance seen in multiphoton ionization \cite{Gribakin1997}. The existence of this initial separation is clear for the static electric field---a similar separation also appears in the present adiabatic case.

\begin{figure}[b]
\includegraphics[scale=0.45,left]{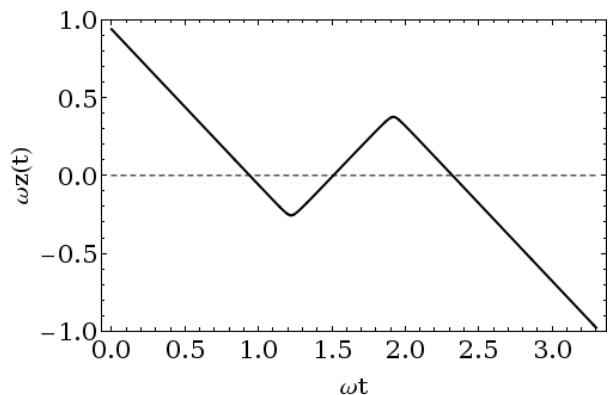}
\caption{The electron follows a classical trajectory $(x_0, y_0, z(t))$ which varies only in the $z$ direction. The $z$ coordinate is shown above for $\xi=10^{5}$. The positron trajectory is given by $-z(t)$. The first collision indicated on the diagram has $n=0$, and is followed by another two collisions with $n=1$ and $n=2$ respectively. The trajectory may potentially intersect with $z(t)=0$ arbitrarily many times, depending on the value of $\xi$, though only the first three collisions $n=0,1,2$ are shown.}
\label{trajectory}
\end{figure}

Since the energy of the $e^-e^+$ collision is large, the fermion annihilation takes place at small separations, $\delta \boldsymbol{r}_{12} \ll \lambda =2\pi/\omega$, where the pair is described by the quantity $\Psi (0,t)$. Incorporating the condition $r_{12}=0$ into the classical trajectory, we observe  firstly that
the trajectory is aligned with the direction of the electric field
\begin{equation}
(r_{12})_i=2 z(t)\,\delta_{i3}, \quad v_i(t)=v_\parallel(t)\,\delta_{i3}~,
\label{d12i}
\end{equation}
otherwise, if the electron and positron had non-zero transverse velocities $v_\perp$ and $-v_\perp$, 
the fermions would drift apart. 
Secondly, the $z$ coordinates of the electron and positron must be equal at some point on the trajectory, which implies
\begin{equation}
z(t)=\int_0^t v(t') \,dt'+{\varepsilon(0)}/{F}=0,
\label{z12}
\end{equation}
where $z(t)$ is the electron $z$ coordinate (when the origin is defined as the midpoint of the electron and positron positions upon creation) and $\varepsilon(0)$ is the initial electron energy. Since $v(t)= v_\parallel(t)$ and $p= p_\parallel$, we henceforth dispense with the subscript $\parallel$ for vectors parallel to the electric field, for simplicity. Figure \ref{trajectory} illustrates such a trajectory along the $z$ axis.
 The lower limit of integration in the time integral signifies that we consider the pair created at $\omega t \approx 0$. Equation \eqref{z12} states that at the moment of time $t>0$ the electron and positron possess the same $z$ coordinate, $z(t)=0$.
Thus, Eqs. \eqref{d12i} and \eqref{z12} describe the classical collision of the two fermions at the moment of time $t$.

The Hessian of $S$, according to \eqref{Spropagate},
equals
\begin{equation}
\frac{\partial^2 S}{\partial p_i\,\partial p_j}=-2\int_0^t\,\frac{\partial^2 \varepsilon(t')}
{\partial p_i\,\partial p_j}\,dt'-\frac{2v(0)}{F} \,\delta_{i3}\delta_{j3},
\label{hessian}
\end{equation}
where $v(0)$ is the initial electron velocity. For the second derivatives of energy we find via Eq. \eqref{d12i}
\begin{equation}
\frac{\partial^2 \varepsilon(t)}
{\partial p_i\,\partial p_j}=\frac {\delta_{ij}-v_i(t)v_j(t)}{\varepsilon(t)}=
\delta_{ij}\bigg\{\begin{array}{cl}
1/\varepsilon(t),        & i=1,2 
\\
m^2/\varepsilon^3(t), & i=3.
\end{array}
\label{d2E}
\end{equation}
The fields we consider are strong enough as to ensure that 
for most of the time $\varepsilon(t)\gg m$. Eq. \eqref{d2E} implies that the region of time over which this inequality holds produces minimal contribution to the integral in Eq. \eqref{hessian}.
However, the electron (positron) trajectory is oscillatory,
such that there exist moments of time (call such a moment $\hat t$) when the
electron (positron) velocity turns zero, making the energy
minimal, $v(\hat{t})=0$, $\varepsilon (\hat{t})=m$.

The vicinities of the mentioned points produce the main contribution to the 
Hessian. Motivated by this observation, we expand the electron momentum 
about such a point $\hat t$,
\begin{equation}
p(t)\approx
(t-\hat t)\frac{dp(\hat t)}{dt}\!=(t-\hat t)F \cos\,\omega\hat t \approx\! \pm (t-\hat t)F.
\label{p(t)=0+deltap}
\end{equation}
Since $p(\hat t)=0$ implies that
$|\sin \omega\hat t|=(\omega/F)|p|\lesssim (\omega/F)(\xi\sqrt F)$
from \eqref{paraperp}; consequently $|\sin \omega\hat t|\lesssim \sqrt F/m\ll 1$, which implies
$|\cos\omega\hat t|\approx 1$.
Using \eqref{p(t)=0+deltap} we obtain the energy close to the moment $\hat t$,
\begin{equation}
\varepsilon^2(t)\approx m^2+(t-\hat t)^2{F^2}.
\label{epsolon^2(t)=1+delta^2}
\end{equation}
The contribution of the point $\hat t$ to the integrals in \eqref{hessian} are
\begin{equation}
\int\frac{dt}{\varepsilon^3(t)}
\approx \int\frac{dt}{\big(m^2+(t-\hat t)^2{F^2} \big)^{3/2}}\approx \frac2{m^2F}.
\label{int epsilon-3}
\end{equation}
Here, we extended the limits of integration to $\pm \infty$, as the area of substantial contribution to the integral
$\delta t\sim m/F$ is much smaller than the period of the oscillation $T=2\pi/\omega$.

Similar to \eqref{int epsilon-3}, we write
\begin{align}
&\int\frac{dt}{\varepsilon(t)}
\approx \int\frac{dt} { \big(m^2+(t-\hat t)^2{F^2} \big)^{1/2}}
\nonumber
\\
&\approx
\frac 1 F \int_{-\pi\xi/2 }^{ \pi\xi/2}\frac{dx}{(1+x^2)^{1/2}}=\frac2F\,\ln\pi\xi~,
\label{int epsilon-1}
\end{align}
where $x=(t-\hat t)F/m$.
By contrast with \eqref{int epsilon-3}, the limits of integration here cannot be extended to infinity
due to the logarithmic divergence at large $x$. We calculate the integral in the large-log approximation presuming that 
$\ln 	\pi\xi \gg 1$. The limits of integration in the last integral in \eqref{int epsilon-1} correspond to the conditions $t_{max}-\hat t=\hat t-t_{min}=T/4$, which we take as the vicinity of the point $\hat t$. We verified numerically that the identity in \eqref{int epsilon-1} holds with accuracy better than 10\%
for the whole range of parameters that comply with our assumptions in Section \ref{assumptions}.

In order to take into account the situation in which $n=0$, we include two additional terms in Eqs. \eqref{int epsilon-1} and \eqref{int epsilon-3}, defining
\begin{align}
Y_\nu(x) &= \int_0^{x} \frac{1}{\left|x-\sin\phi \right|^{\nu}}~d\phi
\label{Yfunctions}
\end{align}

From Eqs. \eqref{hessian}, \eqref{d2E}, \eqref{int epsilon-3} and \eqref{int epsilon-1} we find
\begin{align}
&S_{ij} =\frac{\partial^2 S}{\partial p_i\,\partial p_j} 
\label{Hessian-explicit}=-\frac1F ~\delta_{ij}
\\
&
\times\bigg\{\begin{array}{ll}
2n\,\ln \pi\xi+
\big(1+Y_1(\tfrac{ |p|}{m\xi})\big)\ \delta_{n,0},~ & i=1,2 
\\
2n+v(0)+\xi^{-2} Y_3(\tfrac{ |p|}{m\xi})\ \delta_{n,0},~ & i=3.
\end{array} \nonumber
\end{align}
Above, $n$ is the number of points such as $\hat t$ in the interval $(0,t)$---i.e. the number of points at which the electron and positron
velocities change sign before recollision.

Recalling once again that we have set $r_{12}=0$, we use the saddle point method via 
Eqs. \eqref{wf-saddle-point} and \eqref{Hessian-explicit} to find with accuracy $\mathcal O(\omega/m)$,
\begin{align}
&\Psi(0,t)=\xi_p \, \Phi(0,t)
\label{explicit-wf}
\\
&\Phi(0,t)\approx \sum_{n\ge 0}\frac{F^{3/4}\zeta(p)}{2^{7/4}\,\xi^{1/2} D(p)}
\exp \Big(i (S(p,0,t)-\frac {3\pi}4)\Big)
\nonumber
\\
&D(p)=
\begin{cases}
2n (2n+v(0))^{1/2}\,\ln  \pi\xi, &n\ge 1
\\
(1+Y_1(\tfrac{|p|}{m\xi}))\,(\xi^{-2} Y_3(\tfrac{|p|}{m\xi})+v(0))^{1/2}, 
&n=0
\end{cases}
\label{D}%\nonumber
\end{align}
To write the final expression we calculated $\det(- S_{ij})$ using \eqref{Hessian-explicit},
substituted the constant $A_0$ as well as the function $\zeta$  \eqref{chi} taking into account 
that $p_\perp=0$.

The wave function Eq. \eqref{explicit-wf} is non-zero provided
there exists a classical trajectory, for which both fermions converge at the origin 
for some moment of time $t>0$.

The action $S(p,0,t)$ in \eqref{explicit-wf} is taken on the classical trajectory defined by \eqref{z12}, being equal to
\begin{equation}
S(p,0,t)=-2\int_0^t \varepsilon(t')\,dt'-\frac{2}{F}\int_0^{p}\sqrt{m^2+p'^2} \ dp'~,
\label{Scl}
\end{equation}
as it follows from \eqref{Spropagate} when $\mathbf r_{12}(t)=0$. This action is a function of the collision time, and the initial momentum $p$ (recall $p$ is the initial momentum whilst $p(t)$ is the momentum at time $t$). It follows that the time dependence of \eqref{explicit-wf} is due to the time dependence of $n$ and $p$ which come from requiring $\mathbf r_{12}(t)=0$.

We now verify the applicability of the saddle point approximation---recognizing that it is not immediately clear that we could neglect variations in the function $\zeta$ in Eq. \eqref{chi}. We estimate the typical momenta at which the integrand in \eqref{explicit-wf} is saturated;
according to \eqref{Hessian-explicit} for $n \geq 1$
\begin{equation}
\delta p_\perp^2\, \lesssim \,\frac F{2n \ln \pi \xi},\quad\quad
\delta p_\parallel^2\, \lesssim \, \frac F{2n+1}.
\label{delta p}
\end{equation}
Comparing these estimates with typical momenta defined in \eqref{paraperp}, we find that $\delta p_\perp^2\ll p_\perp^2$,
$\delta p_\parallel^2\ll p_\parallel^2$. This implies that the functions $S$ and $\xi_{\boldsymbol{p}}$ vary slowly for the range of momenta that dominate the integrand \eqref{Phi exp iS}, justifying the saddle point approximation.

The small factor $\xi^{-2}$ 
in front of $p_\parallel^2$ in the exponent of $\zeta$ \eqref{chi} ensures variations in $p_\parallel$ do not vary the integrand dramatically; the second term in the exponent is the function $p_\perp^2/(2F)$, which varies slowly since around the saddle point $p_\perp=0$, 
this function and its first derivative vanish,
while its second derivative,
$1/F$, remains smaller then the second derivatives over $p_\perp$ 
of the action $S$  in \eqref{Hessian-explicit}, $4n \ln(\pi\xi)/F$, which are logarithmically large.
The necessary condition for our argument's validity reads $4n \ln(\pi\xi)\gg 1$, where $n\ge 1$, and is well satisfied for any large $\xi$.

In the situation where $n=0$, the factors in Eq. \eqref{delta p} are replaced by the functions $Y_1(\frac{ p}{\,\,m\xi})$ and $Y_3(\frac{ p}{\,\,m\xi})$. We have verified numerically that for the whole range of parameters considered these functions remain large making the above argument valid.

Summarizing, \eqref{explicit-wf} provides an explicit representation for
the wave function of the $e^-e^+$ pair, $\Psi(0,t)$,
in which the coordinates of the two fermions are equal.

\subsection{Fourier component of the wave function}
\label{Fourier component of wavefunction}

Consider an inelastic collision of the $e^-e^+$ pair that results in the creation of other particles.
We concentrate our attention on the annihilation which produces two photons,
\begin{align}
e^+ + e^-\rightarrow 2\gamma~.
\label{ee2g}
\end{align}
Let the energy of each emitted photon be $\Omega$,
so that the total energy released is $E=2\Omega$.
Following the discussion presented after Eq. (\ref{int dE}) we aim
to calculate the Fourier component $\tilde \Psi (0,2\Omega)$ of the time-dependent wave function $\Psi (0,t)$.
Inverting \eqref{int dE},
\begin{equation}
\tilde \Psi (0,2\Omega)=\int e^{ 2\,i\,\Omega\,t}\,\Psi(0,t)\,dt~.
\label{tilde Psi}
\end{equation}
Our goal is to evaluate this integral explicitly. The task is simplified by the fact that the photon energy
$\Omega$ and fermion energy $\varepsilon$ greatly exceed $\omega$.
We use this fact to execute a second application of the saddle point method.

We note the contributions to rapid variations in \eqref{tilde Psi};
one comes from the factor $\exp{ (2\,i\,\Omega\,t)}$ and another from the wave function \eqref{explicit-wf} which incorporates $\exp{ (i \,S)}$.
Consequently, we find that the fast variation in \eqref{tilde Psi}
is described by the factor $\exp{\left(i \tilde S \right)}$, in which $\tilde S$ is the Legendre transform of $S$, given by
\begin{align}
\tilde S(p,0,2\Omega) &= 2\Omega\,t+S(p,0,t)
\label{cal A}\\
&=2\Omega\,t-2\int_0^t \varepsilon(t')\,dt' -\frac{1}{F}\int_0^{p}\!\sqrt{m^2+p'^2} ~ dp'.
\nonumber
\end{align}
Call $\tau$ the moment of time at the saddle point defined by
\begin{equation}
\partial_t \, \tilde S(p,0,2\Omega) =0\,.
\label{A saddle}
\end{equation}
Then the integral in \eqref{tilde Psi} is simplified to
\begin{equation}
\tilde \Psi (0,2\Omega)=e^{ 2\,i\,\Omega\, \tau}\Psi(0,\tau)\Big(\frac{2\,\pi \,i}{\tilde S''(p_{\tau},0,2\Omega)}\Big)^{1/2}
\label{tilde Psi result}
\end{equation}
where $\tilde S''(p_{\tau},0,2\Omega)$ is the second derivative of $\tilde S(p,0,2\Omega)$ evaluated at the
saddle point (for simplicity, the dependence on $\tau$ is suppressed below, $\tilde S''(p_{\tau},0,2\Omega)\equiv \tilde S''$).

It is worthwhile to reiterate the conditions which the fermions' trajectory must obey.
Firstly, the time coordinate of the saddle point (i.e. the time of recollision) should satisfy Eq. (\ref{A saddle}).
It is shown below in Eq. (\ref{Omega}) that this condition is equivalent to conservation of energy in the annihilation of the electron and positron. 
Secondly, we should keep in mind that the derivation in the previous section required that the pair collide. That is, we are interested in those moments $\tau$ at which, in addition to energy conservation \eqref{A saddle},
\begin{align}
\boldsymbol{r}_{12}(\tau)=0.
\label{z12prime}
\end{align}
This equation creates an implicit relationship between $\tau$ and the initial momentum, which was
initially introduced as a constant---the electron momentum at the moment of creation $p$, 
which we took as $\omega t\approx 0$. We denoted the relation above by subscript. Our task now is to evaluate explicitly the derivatives of $\tilde S$. Since we presume Eq. \eqref{z12prime} valid, we
are to treat $p$ not as a constant, but a function of $t$
defined by this equation, as $t$ is the collision time in \eqref{explicit-wf} as well. 

Hence, before differentiating $\tilde S$, which is a function of $p$, we calculate the derivative $\dot p$.
To this end we differentiate \eqref{z12} finding
\begin{equation}
v(t)+\dot p \int_0^t \frac{\partial v(t')}{\partial p}\,dt'+\dot p\frac {v(0)}F=0~.
\label{v+dot p}
\end{equation}
Since $\partial v(t')/\partial p=m^2/\varepsilon^3(t')$, the integral calculated in \eqref{int epsilon-3} gives an expression for $\dot p$,
\begin{equation}
\dot p=-\frac{v\,F}{2n+v(0)+\xi^{-2}\,Y_3\big(\frac{|p|}{m\xi}\big)\,\delta_{n,0}}.
\label{dot p}
\end{equation}
It is straightforward to show that under Eqs. \eqref{z12} and \eqref{dot p}, the saddle point condition
\eqref{A saddle} implies
\begin{equation}
\Omega=\varepsilon.
\label{Omega}
\end{equation}
Here $\Omega$ is the photon energy and $\varepsilon$ is the
classical energy of the electron at the moment of collision.
Eq. \eqref{Omega} may be considered as the statement of energy conservation---the total energy of the pair at the moment of the collision $2\varepsilon$ goes into the total energy of photons $2\Omega$ after annihilation.
This makes physical sense as the collision is fast, and therefore the external field cannot change the energy of the fermions during the process.

Similar calculations produce
\begin{equation}
\tilde S '' =-2\, F \, v \, w,
\label{ddot A}
\end{equation}
where $v$ is the electron velocity at the moment of collision $\tau$,
while the function $w$ equals
\begin{equation}
w=\cos \omega\tau-\frac{v}{2n+v(0)+\xi^{-2}\,Y_3\big(\frac{|p|}{m\xi}\big)\,\delta_{n,0}}~.
\label{www}
\end{equation}

Combining Eqs. \eqref{explicit-wf}, \eqref{tilde Psi result} and \eqref{ddot A} we find
\begin{equation}
\tilde \Psi (0,2\Omega)
=\xi_p\,\Phi_{2\Omega}(0)\,,
%=K\,\frac{ u^{{(-)}}_\alpha u^{(+)}_\beta}{2\varepsilon}\,,
\label{int Psi dt}
\end{equation}
where

\begin{align}
&\Phi_{2\varepsilon}(0)=-\frac{\pi^{1/2}\,F^{1/4}Q\exp\big(-\frac { p^2}{4\xi^2 F}+i\, \tilde S (p,0,2\Omega)
\big)}{{2^{7/4}\,\xi^{1/2}D(p)}\,v^{1/2}(\tau)}\,.
\label{K}
\end{align}
\begin{align}
&Q=w^{-1/2}(\tau)~.
\label{Q}
\end{align}
This is an essential result. The Fourier component $\tilde \Psi (0,2\Omega)$ equals
the amplitude of the event in which the fermions are located at the same point, i.e. `collide' $r_{12}(\tau)=0$, while their total energy  at the moment of time $\tau$ equals $E=2\Omega$. 
This amplitude is proportional to the simple spinor function $\xi_p$, 
while the coefficient $\Phi_{2\varepsilon}(0)$ takes into account all effects preceding the collision---
the creation of the pair, and its subsequent propagation in the time-dependent field. Since we neglected variations other than those introduced by $\exp (i \tilde S)$, we have
\begin{align}
\Phi_{2\varepsilon}(0)=\int \exp(-2i\,\varepsilon t)\Phi(0,t)\,dt~.
\label{fourier!!!}
\end{align}
We reiterate the physical assumptions incorporated  in (\ref{K}) and the meaning of variables employed there.
Equation (\ref{K}) presumes that there exists a classical trajectory \eqref{z12} for which the initial electron momentum
equals $p=p_\parallel(0)$, while at the moment of time $\tau$ the electron and positron
come to the same coordinate $r_{12}(\tau)=0$, i.e. `collide'. 
Moreover, it is necessary that at the moment of collision the electron energy  $\varepsilon=\varepsilon(\tau)$ satisfy $\varepsilon=\Omega$.
Equation (\ref{K}) explicitly refers to the longitudinal momentum $p=p_\parallel(0)$ of the electron at the moment of creation $t=0$,
and incorporates the corresponding electron velocity $v(0)=v_\parallel(0)$, while the quantity $v=v_\parallel(\tau)$ denotes the electron velocity at the moment of collision.
The integer $n$ marks the number of times the longitudinal electron velocity changes its sign
in the interval $(0,\tau)$; thus $n=0$ indicates the `first collision', i.e. the  event in which $\tau$ takes the smallest possible value when conditions $r_{12}(\tau)=0$ and 
$\varepsilon=\Omega$ are satisfied.

The function $Q$ is written in Eqs. (\ref{K}) and (\ref{Q}) as a separate factor for the following reason:
there exist specific values of the parameters $F,~\xi,~\Omega$ and $n$, for which there is a collision time $\tau^*$ with $w(\tau^*)=0$. Later on, see Eq. (\ref{dWdrho}), we will see that the probability of the annihilation of the $e^-e^+$ pair is $\propto|\Phi_{2\varepsilon}(0)|^2\propto 1/|w|$ and therefore is strongly divergent in the vicinity of such a point.
However, this divergence arises only due to the approximation
made in the integration over time in Eq. (\ref{tilde Psi}). Our calculations relied on the validity of the saddle point method, specifically that $\tau^2 \tilde S '' \gg 1$, which clearly fails when $\tilde S '' \propto w =0$. 

The way out of this problem is well known, see e.g. \cite{Copson2004}. One needs to push the calculation to further accuracy, incorporating the third derivative of the rapidly varying exponent---in our case take into account $\tilde S ''' $ (see Appendix 
\ref{third order}).
The result is that 
in the vicinity of a point where $w(\tau^*)=0$, one needs to amend the absolute value of the $Q$ factor through $|Q|^2 \rightarrow |Q_\text{Ai}(\tau)|^2$, where 
\begin{align}
|Q_\text{Ai}(\tau)|^2 =2\pi X(\tau)\Big(\text{Ai}\big(-X(\tau)~w(\tau)^2 \big) \Big)^2.
\label{airyalt}
\end{align}
Here
\begin{align}
X(\tau)=\Big(\frac{\xi^2|v(\tau)|}{ F |\sin\omega\tau |} \Big)^{2/3}~,
\end{align}
where Ai$(x)$ is a conventional Airy function.
Using its known asymptotes, see e.g. \cite{Copson2004}, one verifies that in the vicinity of the points where $w=0$ the factor $|Q|^2$ is enhanced, but remains finite. In the region well separated from these special points this factor shows rapid oscillations, though its average over these oscillations, 
$\{|Q_\text{Ai}|^2\}$,
satisfies $\{|Q_\text{Ai}|^2\}\approx |Q|^2=1/|w|$. 

This situation is similar to the one encountered in optics in a vicinity of a caustic. The distinction is that instead of the 
interference originating from different regions of the wave-front, as is the case in optical caustic, 
we are facing the interference in the integral in (\ref{tilde Psi}) originating from contributions of different moments of time. Recently, similar peaks were observed in the frequency spectrum of Compton scatting of electrons by a laser pulse \cite{Seipt2015}.

Another useful way to look at the physical meaning of Eq. (\ref{airyalt}) is to consider the 
time dependent energy of the pair $2\varepsilon(t)$. As we know, the annihilation takes place
at the point $\tau$ where $2\varepsilon(\tau)=2\Omega$. Clearly, if at the same moment of time $\frac{d\varepsilon(t)}{dt}=0$, the probability of the transition is enhanced because the transition has the opportunity  
`to last longer' compared with the generic case when
$\frac{d\varepsilon(t)}{dt}\ne 0$. 
%Put otherwise, at these points $\tau^*$, the collision energy is stationary with respect to variations of initial energy, so the number of collisions with energy $\varepsilon(\tau^*)$ is enhanced. 
This situation closely resembles conventional adiabatic transitions between different time dependent energy terms, for example the Landau-Zener transition, see e.g. \cite{Landau1977}.

Summarizing, Eqs. \eqref{int Psi dt}$-$\eqref{airyalt} give an explicit representation for  
$\Psi(0,2\Omega)$, the Fourier component of the wave function of the $e^-e^+$ pair taken at the small separation between the   fermions.

Our treatment of the semiclassical problem
relies on the saddle point approximation.
A possible alternative would have been to work in the van Vleck determinant formalism \cite{VanVleck1928}.
An advantage to our method we employ is that it efficiently overcomes the problem of caustics, using the approach standard in WKB theory. In fact, ignoring caustics, our saddle point calculation is equivalent to the van Vleck determinant, see Appendix \ref{VVD}.

The creation amplitude is the same for all values of $y$ and $z$. The pair creation therefore occurs coherently at all points in the $yz$ plane. Note the only stipulations made in deriving \eqref{int Psi dt} were $\nabla_{\boldsymbol{p}}\,S=0$, $\partial_{t}\,\tilde S=0$ and $r_{12}=0$. We do not assume the $e^-e^+$ collision occurs at $\omega z =0$, as $z(t)$ is the relative separation of the fermions --- there is no unique origin $\omega z =0$ in this problem. The quantity \eqref{int Psi dt} therefore represents the total probability density of all collisions at all points in the laser focus along the $yz$ plane, due to the fermions created at $\omega t \approx \omega x \approx 0$.

\section{Radiation reaction}
\label{rad}

The wave function \eqref{K} gives a semiclassical description of the $e^-e^+$ pair. The corrections to this formula should be of order $\mathcal O (\omega/m)$, and so long as $\omega \ll m$ the semiclassical description remains valid. However, as demonstrated in \eqref{paraperp}, the motion of the $e^-e^+$ is relativistic and one should therefore consider possible corrections due to the photon emission --- which may cause the electron or positron to radiate and recoil during their oscillations in the laser field. 

However, Eq. \eqref{d12i} shows that the created fermions move predominantly linearly along the direction of the electric field. 
%This simple and important fact 
%implies that in the problem at hand the photon radiation is suppressed.
This distinguishes the present situation from the general case. 
When the electron interacts with a plane wave
the radiated energy may be large \cite{Nikishov1964}, with dramatic implications such as the production of $e^-e^+$ and $\gamma$ cascades \cite{Narozhnyi2015, Fedotov2010}, as well as 
other classically unexpected phenomena \cite{Nerush2011, Bell2008, DiPiazza2010}. 
Recall also that the radiation spectrum of the fermion propagating in a general electromagnetic field can be approximated by that in a plane wave, and play a similarly crucial role, provided the parameters of the electromagnetic field satisfy 
\begin{align}
\chi^2 \ge 1, \quad  \chi^2 \gg \mathcal F, \mathcal G
\label{chiFG}
\end{align}
where 
$\chi^2 = {e^2(F_{\mu\nu} p_\nu)^2}/{m^6}$,
$\mathcal F= {e^2F_{\mu\nu}^2}/{m^4}$,
and $\mathcal G = {e^2 F^*_{\mu\nu}F_{\mu\nu}}/{m^4}$, see \cite{Nikishov1964, Fedotov2010}. The point is that in the problem considered here, when the fermions propagate along the direction of the electric field, both conditions in (\ref{chiFG}) are violated.
%Recall that the component of an electric field longitudinal to the direction of an observer's motion is not boosted --- only those components of the field transverse to the direction of motion acquire the relativistic factor $\gamma=(1-v^2)^{-1/2}$. Hence, for linear motion along an electric field we expect there there to be no enhancement of the electric field and therefore the quantum electrodynamics are essentially the same in the laboratory and $e^\pm$ reference frames.
%Explicitly, since
%\begin{align}
%\frac{dp_\mu}{d\tau} = F_{\mu\nu} u^\nu
%\end{align}
%it follows that
Indeed, straightforward calculations show
\begin{align}
\chi^2 =
%&= \frac{1}{m^2}\frac{dp^\mu}{d\tau}\frac{dp_\mu}{d\tau} \nonumber 
\frac{e^2 \big(
(\boldsymbol{E}+\boldsymbol{v}\times\boldsymbol{B})^2-(\boldsymbol{v}\cdot \boldsymbol{E})^2
\big) }{m^4(1-v^2)}
= \left(\frac{e\mathcal E}{m^2}\right)^2,
\label{para}
\end{align}
where the last identity relies on $\boldsymbol{v}$ being parallel to $\boldsymbol{E}$, and $B\approx 0$.
We find therefore that in our case $\chi^2=\mathcal F=(F/m^2)^2\ll 1$, which contradicts Eq. (\ref{chiFG}) and hence makes the argument of Ref. \cite{Nikishov1964} inapplicable; 
one cannot rely on the plane wave formulae of Nikishov and Ritus.

The fact that $\chi^2 \ll 1$ makes it plausible that classically forbidden radiation does not play a substantial role; this intuition is corroborated by our full-scale quantum calculations to be presented elsewhere. 
On the other hand, addressing the problem classically one recalls that for a charge moving linearly along the electric field, the radiation of high frequency photons is suppressed unless 
$\mathcal E \ll \alpha^{-1}\mathcal E_c$ \cite{Jackson1998}. Since we assumed that $F lesssim m^2$, the classical radiation is not significant either. 

Thus, the linear kinematics of a $e^-e^+$ pair created in a pure electric field implies that the pair mostly retains the energy it gains during its wiggling in the field. We compare this with \cite{Bulanov2010}, who found the radiation reaction plays a significant role in a circularly polarized laser beam, but less so in a linearly polarized one. The situation changes only when the pair collides, as is outlined in Section \ref{totalprob}.

\section{Probability of annihilation}
\label{totalprob}

Since the inelastic collision of the $e^-e^+$ pair occurs over a short time interval, the factor $\Phi_{2\varepsilon}(0)$ in the wave function $\tilde \Psi (0,2\Omega)$ in (\ref{int Psi dt}) can be considered as 
a constant coefficient, while $\xi_p$, which
incorporates two Dirac spinors, can be considered as a constant spinor function for the two colliding plane waves. As a result, 
the matrix element describing the collision can be factorized, written as
$\Phi_{2\varepsilon}(0) \,M(\varepsilon)$, where $M$ is the matrix element describing the collision of the two fermions in vacuo. 
Correspondingly, the probability $W$ of the inelastic collision, reads
\begin{equation}
dW=|\Phi_{2\varepsilon}(0)|^2\,\Big ({\sum}^\prime |M(\varepsilon)|^2\,\frac{d\rho}{d\varepsilon}\Big)d\varepsilon.
\label{dWdrho}
\end{equation}
Specifically, we continue to examine the scenario in which there are two photons in the final state, as stated in \eqref{ee2g}.
In Eq. \eqref{dWdrho}, $d\rho$ refers to the conventional density of final states for these photons, the summation ${\sum}^\prime$ runs over all possible quantum numbers in the final state, except for the energy of particles $\varepsilon=\Omega$.

To clarify the meaning of \eqref{dWdrho}, let us write the expression for the cross section of the annihilation $e^+e^-\rightarrow 2\gamma$ in the vacuum when the photon energy equals $\Omega$
\begin{equation}
\sigma(\Omega)=\frac{2\pi}{v_{12}}\int d\varepsilon \,{\sum}^\prime \delta(2\varepsilon-2\Omega)|M(\varepsilon)|^2\frac{d\rho}{d \varepsilon}.
%=\frac{\pi}{2v}\,\sum |M(\varepsilon)|^2\,\frac{d\rho}{d\varepsilon}
\label{dsigma}
\end{equation}
Here $v_{12}=2|v|$ is the flux of the colliding fermions for the
chosen normalization of spinors \eqref{uu}.
%and $\int\sum$ means the summation over all
%quantum numbers in the final states, continuous and discrete.
The necessary for us matrix element $M(\varepsilon)$ can
be extracted from Eq.(\ref{dsigma}) and rewritten in terms of the annihilation cross section and the electron velocity
\begin{align}
{\sum}^\prime |M(\varepsilon)|^2 \,\frac{d\rho}{d\varepsilon}=\frac 2 \pi\,|v|\,\sigma(\varepsilon). 
\label{M2sigma}
\end{align}
(Remember that a similar approach is conventionally used for the annihilation of positronium.)
Substituting (\ref{M2sigma}) into \eqref{dWdrho}, and remembering \eqref{K}, 
we find the following expression for the spectral 
distribution of the photons emitted due to the annihilation of one $e^-e^+$ pair
%\begin{align}
%\frac{dW}{d\varepsilon}=
%\frac{\pi}{2^{9/2}}~
%\frac{m\gamma}
%{(\ln \frac \pi\gamma)^2}\left(\frac{F}{m^2}\right)^{1/2}\sum_{n\ge 2}
%\frac{\sigma(\varepsilon)~\exp\big(-\frac {\pi\gamma^2 p^2}{2F}\big)}{n^2\,(2n+v(0))\,|w(\tau)|}~.
%\label{dWdeps}
%\end{align}
\begin{align}
\frac{dW}{d\varepsilon}&=
\frac2\pi\,|\Phi_{2\varepsilon}(0)|^2\,|v|\,\sigma(\varepsilon)
\nonumber
\\
&=
\frac{{F}^{1/2}\,\sigma(\varepsilon) }{4\sqrt{2}\xi}
\sum_{n\ge 0}\,
\frac{\,|Q|^2}
{|D(p)|^2}\,\exp\big(\!-\frac {\pi p^2}{2\xi^2F}\,\big)
\label{dWdeps}
\end{align}
We took into account here that collisions distinguished by the number of oscillations $n$ are separated be large intervals of time $\delta t\propto 1/\omega$. Correspondingly, the phase factors, which appear
in the wave function for the $e^-e^+$ pair, differ very strongly for different $n$.
As a result the interference of contributions from different $n$ to the wave function is suppressed and hence neglected in (\ref{dWdeps}).

To elucidate the physical meaning behind (\ref{dWdeps}), it is convenient to present 
it more explicitly using Eqs. (\ref{D}) and (\ref{Q}) for the factors $D(p)$ and $Q$.
To simplify consideration, we restrict our discussion here to the 
case $n\geq 1$, when we find
\begin{align}
\frac{dW}{d\varepsilon}=
\frac{\pi}{2^{9/2}}~
\frac{m}
{\xi\ln^2 \pi\xi}\left(\frac{F}{m^2_e}\right)^{1/2}\sum_{n\ge 1}
\frac{\sigma(\varepsilon)~\exp\big(-\frac {\pi p^2}{2\xi^2F}\big)}{n^2\,(2n+v(0))\,|w(\tau)|}~,
\label{dWdeps2}
\end{align}
Each aspect of this formula can be easily comprehended. The exponent comes from the momentum distribution of $e^-e^+$ pairs found in Section III. The approximate $\propto n^{-3}$ dependence represents quantum diffusion of the pair. 
The presence of the cross-section gives the probability that the pair collides, 
whilst the prefactor outside the summation is a kinematic factor due to the pair moving along the classical trajectory. The classical trajectory serves as a basis for the semiclassical treatment, so that Eq. (\ref{dWdeps}) gives a proper quantum description of the whole process. 
%This latter comment is essential because, as we shall see, the quantum result greatly 
%surpasses estimations based on simple classical arguments.
Returning to Eq. (\ref{dWdeps}) note that the factor $Q$ there can be taken in the simple form of Eq. (\ref{Q}) only provided $w(\tau)$ is not close to zero. In the vicinity of the singular point, where $w(\tau)= 0$, the more sophisticated expression in (\ref{airyalt}) should be used. 

The total probability $\Upsilon$ that a created $e^-e^+$ pair recollide, 
which we shall call the conversion coefficient, equals
\begin{align}
{\Upsilon}=\int_{m}^{\varepsilon_\text{max}}\frac{dW}{d\varepsilon}\,d\varepsilon~.
\label{conversion}
\end{align}
This probability depends on the electric field and, generally speaking,
should be averaged over the field variation the same way we averaged 
the probability of the $e^-e^+$ production in Section \ref{distributions}, 
but we will not dwell on these details here. The upper limit of integration $\varepsilon_\text{max}$ 
in (\ref{conversion}) is the maximum energy of recollision, which approximately equals $\approx m\xi$, though strictly speaking is different for different values of 
$n$, as shall be investigated in more detail in Section \ref{Calculations}.

Up to now, we discussed the annihilation into two photons.
With small modifications, the general formalism developed in Eq. \eqref{dWdeps}
can be applied to any process $e^++e^-\rightarrow X_1+X_2+\dots$, where 
$X_k$ are some neutral particles. For example, one can similarly consider
the three-photon annihilation substituting the cross section for the three-photon process into \eqref{dWdeps} . 
For production of massive particles $X_k$, the energy $\varepsilon$ needs to be large enough
to allow the process to proceed. 

A more substantial modification is necessary if
there are charged particles in the final state.
The photon energy $\Omega$ does not depend on time or the direction of the photon propagation. By contrast, for the creation of a charged particle, one needs to consider its time dependent energy $E_k(t)$ due to the laser field; which should be taken at the moment of collision $t=\tau$ defined by $2\varepsilon(\tau)=\sum_k E_k(\tau)$, c.f. (\ref{Omega}). Importantly, the term $E_k(t)$ depends on the particle momentum. As a result, the
integration over momenta of this particle in the final state cannot readily be fulfilled analytically.

Summarizing, Eq. \eqref{conversion} gives the probability 
of the creation of two photons per one created $e^-e^+$ pair,
while \eqref{dWdeps} describes the photon spectrum.

 \section{Analytic estimate}
\label{Analytic estimation}
A simple estimation for the conversion coefficient $\Upsilon$ can be made
via classical arguments. The $e^+e^-\rightarrow 2\gamma$ cross section
is of the order of $\sigma\sim \pi\,r_e^2$, see (\ref{Page-cs}). At the same time
typical separations of the fermion pair are $R\sim v/\omega\sim 1/\omega$ because the fermion velocity is $v\approx 1$. 
These arguments lead to the following estimate for the conversion coefficient
\begin{align}
\Upsilon_\text{class}\,\sim\,\frac{\sigma}{R^2}\,\propto  \,\frac{\omega^2 \sigma}{c^2}~.
%\,=\,\frac{F^2 r_e^2 }{m^2\xi^2}~
\label{Ups-class}
\end{align}
(we shall, for the moment, work in absolute units). However, this result grossly underestimates the effect.
One of the reasons is that Eq. (\ref{Ups-class}) does not take into account 
the alignment of the $e^-$ and $ e^+$ velocities along the electric field.
Another is that (\ref{Ups-class}) neglects quantum effects, which prove to be essential.
This is an unexpected finding. At first glance, it seems that the classical 
description in a slow varying laser field should be adequate, at least qualitatively,
but it is not.

To investigate $\Upsilon$ more accurately, take the first line from Eq. (\ref{dWdeps}) and rewrite the conversion coefficient from (\ref{conversion}) as
\begin{align}
\Upsilon=4\langle v\sigma \rangle\int|\Phi_{2\varepsilon}(0)|^2\,\frac{d\varepsilon}{2\pi}.
\label{averaged s}
\end{align}
This equality can be considered the definition of the averaged value 
$\langle v\sigma \rangle$ for the quantity $ v \ \sigma(\varepsilon)$. 
Using Parseval's theorem, we can rewrite (\ref{averaged s}) further
\begin{align}
\Upsilon=2\,\langle v\sigma \rangle \int| \Phi(0, t)|^2\,dt.
\label{averaged s0}
\end{align}
Presuming that the collisions which take place at different $n$ do not interfere with each other, as they are separated by large intervals of time, we also neglect the term with $n=0$, as its contribution to $\Upsilon$ is smaller than those with $n\neq0$.
As a result, from (\ref{explicit-wf}), we develop (\ref{averaged s0}) as
\begin{align}
\Upsilon=\frac{F^{3/2}\,
\langle v\sigma \rangle }{2^{5/2}\xi}
\sum_{n\ge 0}
\int
\exp\Big(\!-\frac {\pi p^2}{2\xi^2 F}\Big)\,
\frac{dt}{|D|^2}.
\label{averaged s1}
\end{align}
Rewriting the integration over the moment of 
collision $t$ via the initial momentum $p$,
$dt=dp/|\dot p|$.
Substituting $\dot p$ from (\ref{dot p}) and recalling (\ref{D}),
we find
\begin{align}
\Upsilon \,&\lesssim \,\frac{F^{1/2}\,
\langle v\sigma \rangle }{2^{9/2}\,\xi\ln^2(\pi\xi)}
\,\sum_{n\ge 1}\frac1{n^2}\,
\int_0^\infty
\exp\Big(\!-\frac {\pi p^2}{2\xi^2F}\Big)\,\frac{dp}{|v|}
\nonumber
\\
&\,\lesssim
\frac{\pi^2}{192}\,\frac{c^3}{\hbar}\frac{F}{\ln^2(\pi\xi)}\,\langle v\sigma \rangle 
\label{averaged s2}
\end{align}
In the last transition in (\ref{averaged s2}) we presumed that the velocity at the moment of collision is large, $|v|\approx 1$, which is usually an adequate approximation.

An interesting aspect of \eqref{averaged s2} is that, excepting the $\xi$-dependence in $\sigma$, this formula has subleading dependence on $1/\xi$. Whilst tending to zero in the static limit, $\Upsilon$ tends to zero very slowly, as $\propto 1/\ln^2(\pi\xi)$.

To simplify our estimate further we can use Eq. (\ref{Page-cs}) to state that
$\langle v \sigma \rangle \approx \pi r_e^2 (m/ \tilde \varepsilon )^2$,
where $\tilde\varepsilon$ is the typical energy of the collision, which is expected during the annihilation for the given parameters of the laser field. This gives
\begin{align}
\Upsilon_\text{quant} \,\lesssim \,
\frac{\pi^3}{192}\,\frac{c^3}{\hbar}\,\frac{r_e^2\,F}{\ln^2(\pi\xi)} \left(\frac{m}
{{\tilde\varepsilon}}\right)^2
\label{Ups-quant}
\end{align}
We used here the natural units to emphasize a notable fact --- that the estimation found incorporates the Planck constant, in accord with its quantum nature.

For the ratio of the quantum \eqref{Ups-quant} and classical (\ref{Ups-class}) results we find
\begin{align}
\frac{\Upsilon_\text{quant}}{\Upsilon_\text{clas}}\propto
\frac{c^3}{\hbar}\,\frac{\xi^2}{ \ln^2(\pi\xi)}
\frac{m^2}{F} \gg 1~.
\label{q/cl}
\end{align}
where in the classical approximation we estimated 
$\langle \sigma \rangle \approx \pi r_e^2 (m/ \tilde \varepsilon)^2$, as in the quantum approach. 
Importantly, Eq. (\ref{q/cl}) states
that the estimate based on proper quantum description greatly surpasses
expectations using classical parameters of the problem (remember that $\xi \gg 1$ and $F\ll m^2c^3/\hbar$). 

It is also interesting that Eq. (\ref{Ups-quant}) gives an upper limit for the conversion coefficient,
\begin{align}
\Upsilon\le \Upsilon_\text{max}=\frac{\pi^3}{192}\,\frac{r_e^2\,F}{\ln^2(\pi\xi)}=
\frac{\pi^3}{192}\,\frac{\alpha^2\,F}{m^2\ln^2(\pi\xi)}.
\end{align}
Another useful application of  (\ref{Ups-quant}) arises if we take into account
that $\tilde\varepsilon$ should grow with the $\xi$ increase,
$\tilde\varepsilon\propto m\xi$. This implies, from Eq. \eqref{averaged s2}, that
\begin{align}
\Upsilon_\text{quant} \,\approx\, \frac{\pi^3}{192} \frac{\alpha^2\,F}{\xi^2\ln^2(\pi\xi)}.
\label{Ups-quant-g=0}
\end{align}
In the following section we note that for the $n=1$ case, $\tilde\varepsilon$ exhibits weaker growth with $\xi$,
$\tilde\varepsilon\propto m\xi^{1/4}$. This implies
\begin{align}
\Upsilon_\text{quant} \,\approx\, \frac{\pi^3}{192}\frac{\alpha^2\,F}{\xi^{1/2}\ln^2(\pi\xi)}.
\label{Ups-quant-g=0-n=1}
\end{align}
The formula \eqref{Ups-quant-g=0-n=1} is compared with numerics in Figure \ref{estimation} of Section \ref{discussion}.
In the static case $\xi\rightarrow \infty$, the conversion coefficient equals zero---no photons are produced as the static field cannot recollide created pairs \cite{Kuchiev1987}.

\section{Photon production}
\label{Calculations}
Here we present numerical calculations of the conversion coefficient and the spectrum of emitted high energy photons, based on Eqs. (\ref{dWdeps}) and 
(\ref{conversion}).
It is convenient to investigate particular values of $n$ separately.

The calculations are organized as follows.
One fixes values the parameters $\xi$ and $F/m^2$. 
As explained previously, calculation of the spectrum requires the classical trajectory of the electron in the oscillating electric field $F\cos\omega t$. 
We take the initial longitudinal electron velocity $v(0)$; the corresponding initial energy $\varepsilon(0)$ gives the initial separation between the electron and positron, $z(0)=2\varepsilon(0)/F$. With these initial conditions, the equation of motion is integrated. Then we choose those $v(0)$ 
for which there are $n$ moments of time $\tau_n>0$ with  $z(\tau_n)=0$. 
%We exclude initial conditions such that there do not exist $n$ such times $\tau_n$. 
%The electron energy at each such moment of time gives the corresponding photon energy, $\Omega=\varepsilon(\tau_n)$. 

We also require the cross section for the $e^+e^-\rightarrow 2\gamma$ collision in the center of mass frame of the $e^-e^+$ pair in the case where the projection of the total spin
on the pair momentum is zero. 
According to Page \cite{Page1957}, this cross section reads
\footnote{Ref. \cite{Page1957} introduces two cross sections, $\sigma_1$ and $\sigma_2$, from which the cross sections for different polarizations of the fermion pair can be constructed. 
Simple analyses shows that in our case it is 
necessary to take $\sigma=2\sigma_1$, which is done in Eq. (\ref{Page-cs}).}
\begin{align}
\sigma(\varepsilon)=\pi \,r_e^2\,\frac{1-v^2}{2v}\,&\bigg(
\frac{(1-v^2)(2v^4+3v^2-3)}{2v^3}\ln\frac{1+v}{1-v}
\nonumber
\\
&+\frac{3-3v^2+2v^4}{v^2}\bigg)~.
\label{Page-cs}
\end{align}
Here $\varepsilon$, $v$ and $r_e$ are the electron
energy, velocity and classical radius.
The asymptotic behavior of the cross section reads
\begin{align}
\sigma(\varepsilon)\rightarrow
\pi \,r_e^2\,
\begin{cases}
1/v, &v\rightarrow 0,
\\
m^2/\varepsilon^2, &v\rightarrow 1,
\end{cases}
\end{align}
and, for the whole range of energies, the inequality holds
\begin{align}
\pi r_e^2\,\le \, v \,\sigma(\varepsilon)\,.
\label{ineq}
\end{align}
Solving $z(\tau_n)=0$ gives us the collision times $\tau$ and in turn the collision energies
%i. e. the energy with which the fermions recollide is 
\begin{align}
\varepsilon(\tau_n) = m\xi \Big(\xi^{-2}+\Big(\frac{p}{m\xi}+\sin\omega\tau_n\Big)^2\Big)^{1/2}
\label{energy}
\end{align}
We verified that for most values of $n$ the typical energies are of the order of $m\xi$. 
When the adiabatic parameter is large $\xi \gtrsim 10^{2}$ this energy is sufficient to produce muons, and when it is very large $\xi\gtrsim 10^{6}$ the collision energies may be as large as TeV. The cross section \eqref{Page-cs} contains a factor of $1/v$, meaning slow collisions may play a significant role, producing an enhancement of photon production at energies $\Omega\sim m$. We find that for $n=1$ these slow collisions dominate, but for all other $n$ the high energy collisions contribute either comparably or dominantly.

\begin{figure}[htp]
\flushleft
\vspace{0.2cm}
\subfloat[Differential creation rate for $n=1$ and $\xi=10^{}$]
{
\includegraphics[width=0.45\textwidth]{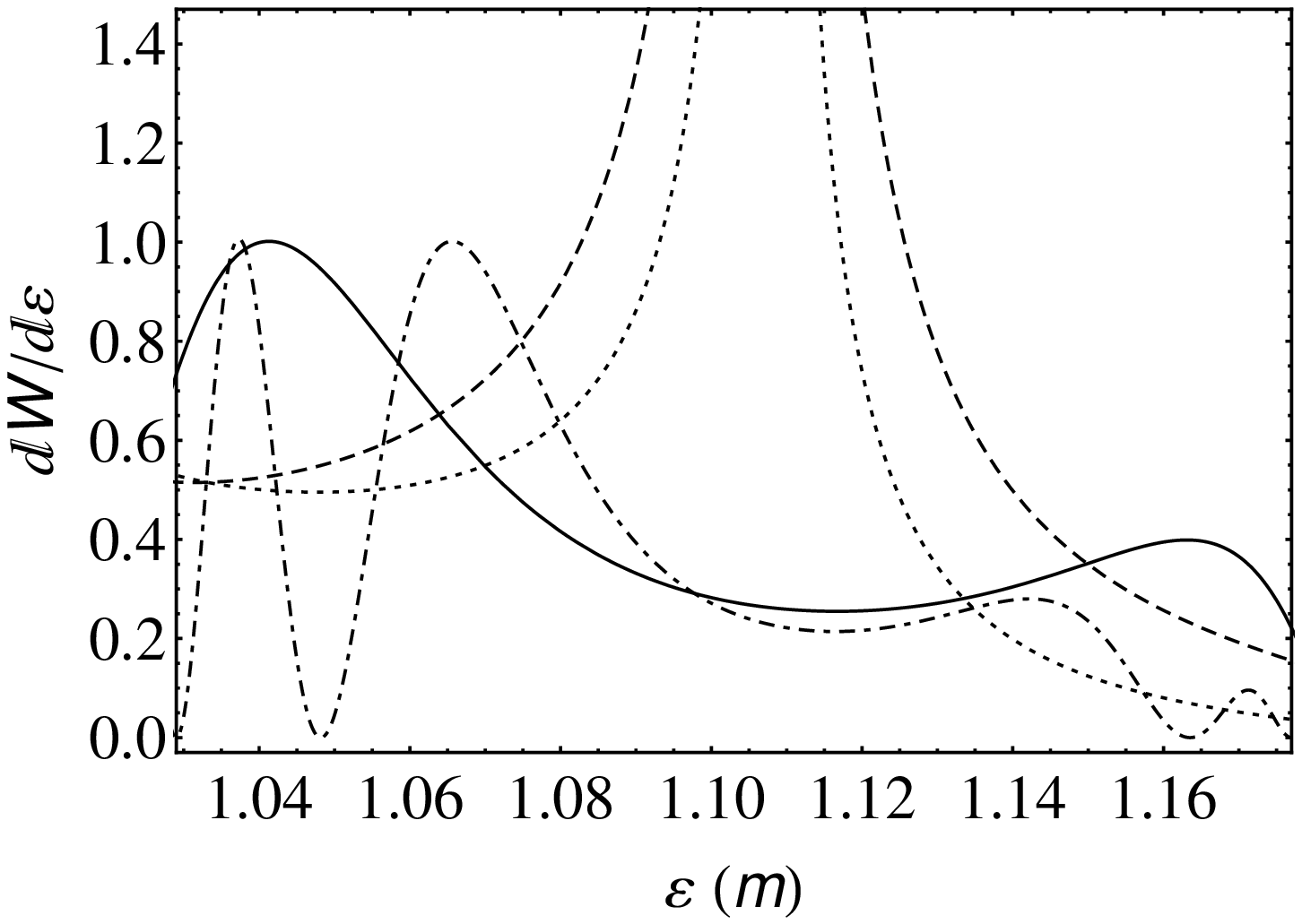}
} \hspace{0.1cm} 
\subfloat[Differential creation rate for $n=1$ and $\xi=10^{4}$]
{
\includegraphics[scale=0.43]{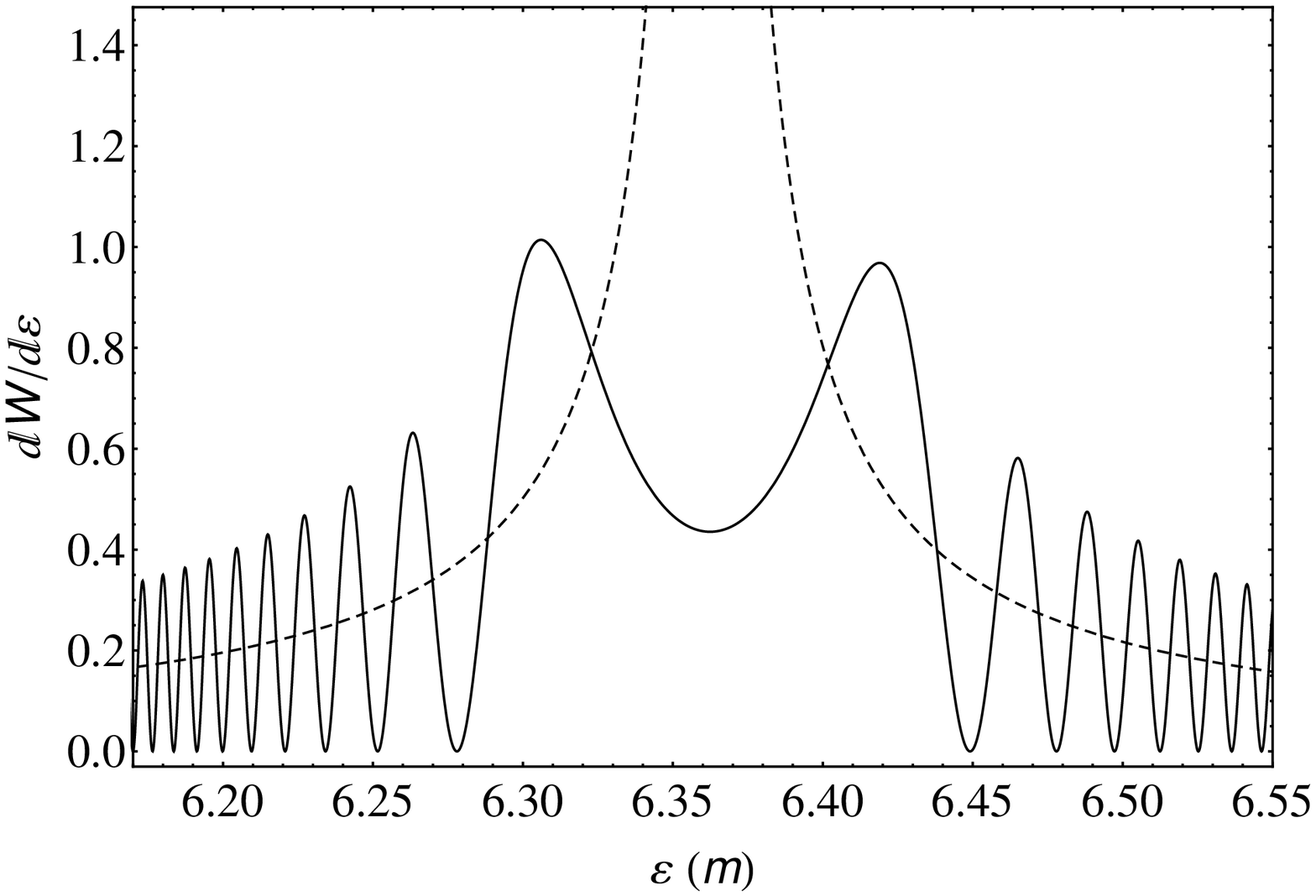}
} 
 \hspace{0.1cm} 
\subfloat[Differential creation rate for $n=1$ and $\xi=10^{6.5}$]
{
\includegraphics[scale=0.63]{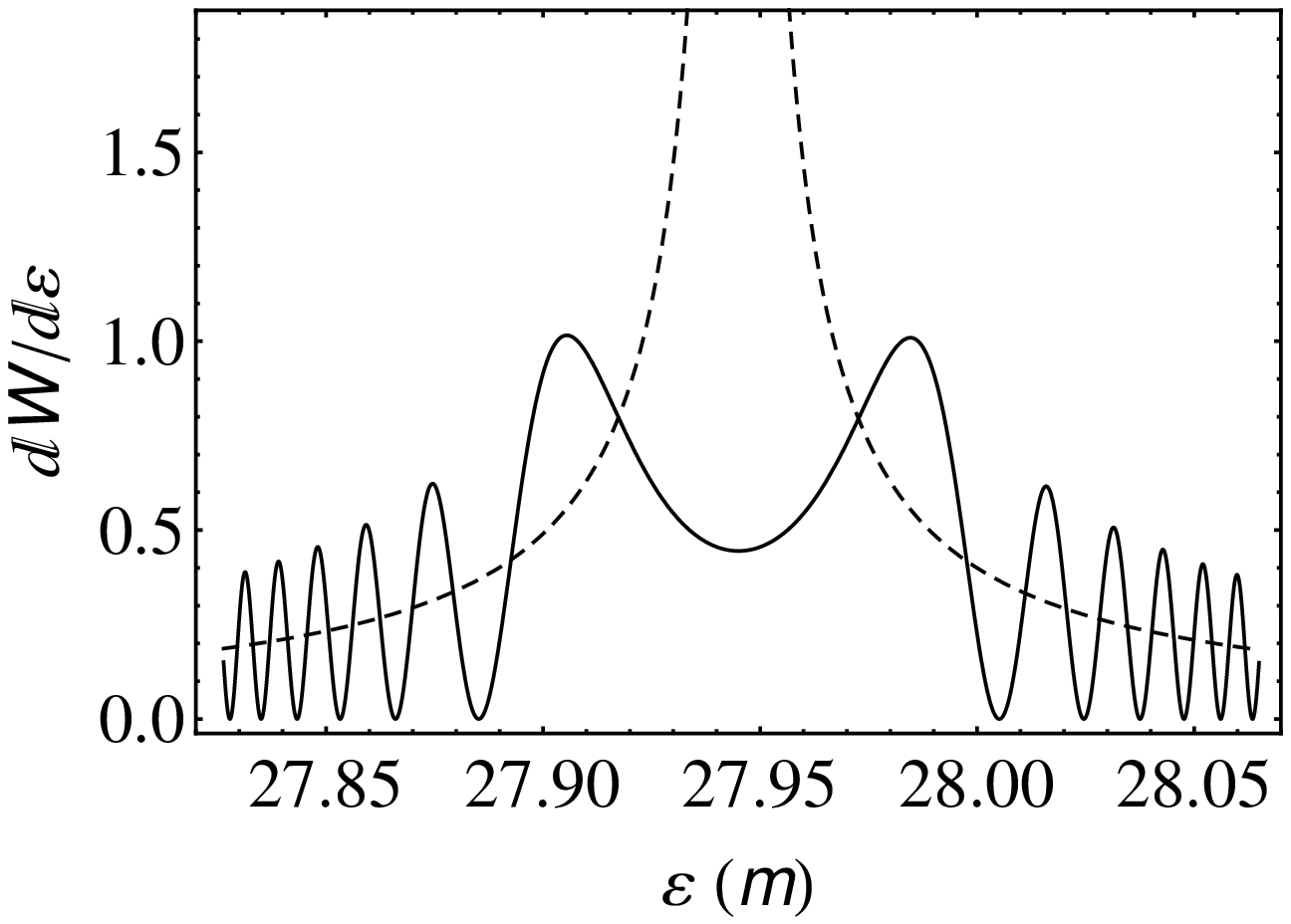}
}
\caption{Photon differential creation rate \eqref{dWdeps} (arbitrary units) versus photon energy (electron mass) for the process $e^-e^+\rightarrow 2\gamma$, with $n=1$. The dashed (and dotted in (a)), curves represent the creation rate in Eq. \eqref{Q}, whilst the solid (dot-dashed) oscillations represent the amended Eq. \eqref{airyalt}; the latter averages to the former and is finite where the former diverges. Parameters: (a) $\xi=10$, $F/m^2~=~0.1$ (dot-dashed, dotted) $F/m^2~=~0.4$ (solid, dashed), (b) $F/m^2~=~0.1$, $\xi=10^{4}$, and (c) $F/m^2~=~0.1$, $\xi=10^{6.5}$.}
 \label{creationraten1}
\end{figure}

Figures \ref{creationraten1} and \ref{creationraten0} show our numerical results:
the photon spectrum (\ref{dWdeps}) for collisions taking place at $n=1$ and $n=0$ respectively.
The corresponding values of $\Upsilon$ are depicted
in Fig \ref{upsilonnumeric}. Note: Log denotes Log$_{10}$.

\begin{figure}[htp]
\flushleft
\vspace{0.2cm}
{
\includegraphics[scale=0.60]{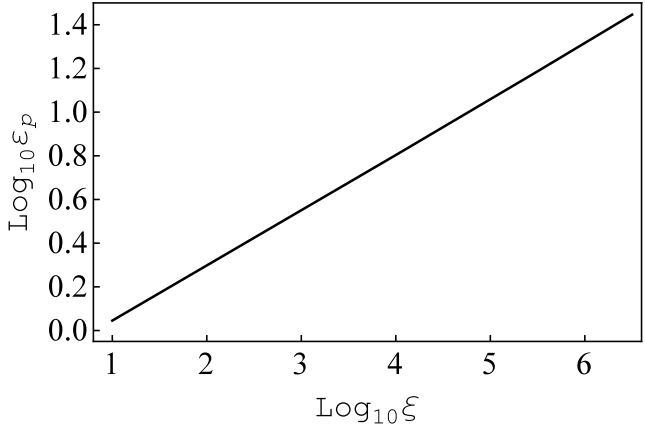}
}
\caption{Peak $n=1$ energy $\varepsilon_p$ versus $\xi$ (log-log scales). The linear relationship shows $\varepsilon_p$ exhibits power-law $\xi$ dependence.}
 \label{epeak}
\end{figure}

\begin{figure}[htp]
\flushleft
\vspace{0.2cm}
\hspace*{-0.1cm}
\subfloat[Rescaled $n=0$ creation rate for $\xi=10^{1.1}$]{
\includegraphics[scale=0.53]{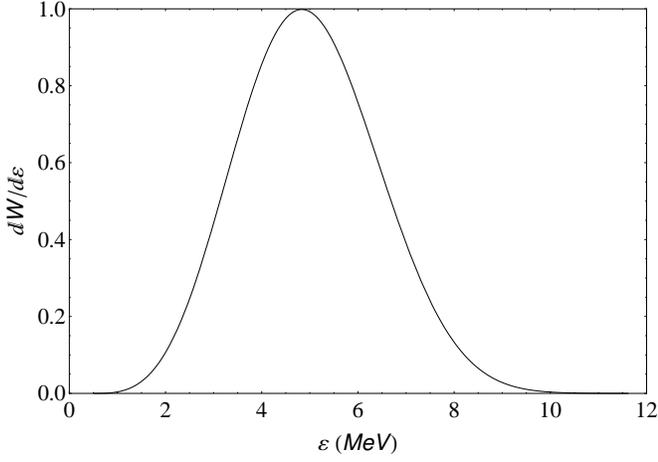}
}  \hspace{0.1cm} \subfloat[Rescaled $n=0$ creation rate for $\xi=10^{4}$]{
\includegraphics[scale=0.53]{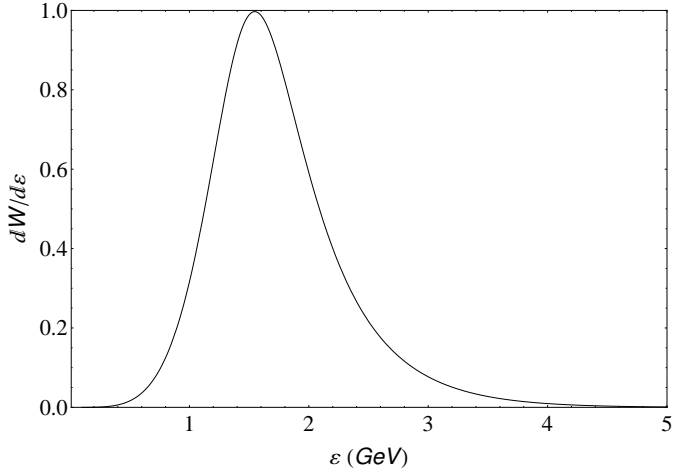}
} 
\caption{The photon spectrum \eqref{dWdeps} (arbitrary units) versus photon energy for the first ($n=0$) $e^-+e^+\rightarrow 2\gamma$ collision. The electric field is $F/m^2~=~0.1$, (a) $\xi=10^{1.1}$ (b) $\xi=10^{4}$.}
 \label{creationraten0}
\end{figure}

\begin{figure}[h]
\flushleft
\vspace{0.8cm}
\subfloat[Log-log plot of $\Upsilon$ against $\xi$, for $n=0$]{
\includegraphics[width=0.45\textwidth]{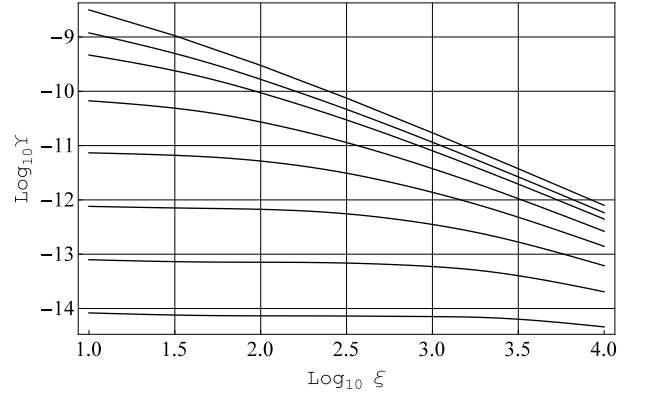}
} \hspace{0.4cm} \subfloat[Log-log plot of $\Upsilon$ against $\xi$, for $n=1$]{
\includegraphics[width=0.45\textwidth]{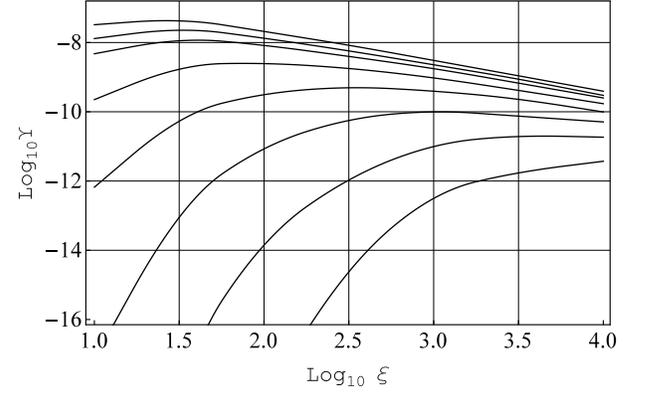}
} \\
\subfloat[Log-log plot of $\Upsilon$ against $\xi$, for $n=0,1$]{
\includegraphics[width=0.45\textwidth]{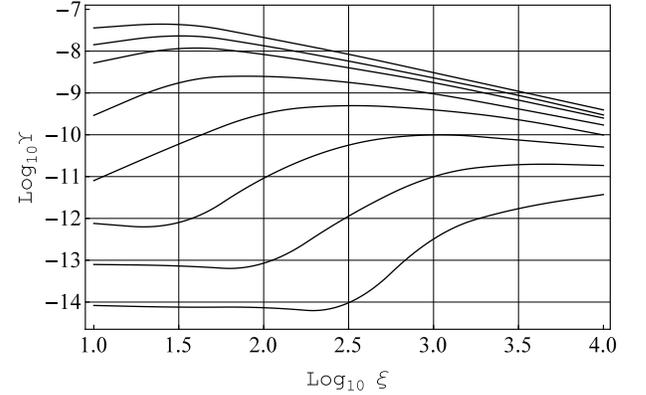}
}
\caption{Log-log plots of numerical calculations for $\Upsilon$; (a) $n=0$, (b) $n=1$ and (c) total $n=0,1$. The curves correspond to $F/m^2~=~0.0125, ~0.025, ~0.05,~ 0.1, ~0.2, ~0.4, ~0.6, ~1$ (bottom to top).}
\label{upsilonnumeric}
\end{figure}

Observe that the spectra and the conversion coefficients exhibit qualitatively different behavior for the cases of $n=0$ and $n=1$ collisions. 
The $n=0$ collisions of Fig. \ref{creationraten0} show broad spectrum widths which incorporate high energies 
as the collision energy scales $\propto\xi$. 
By contrast, the spectrum for $n=1$ is sharp, and occurs at much lower energies. 
The typical energy released scales roughly as $\propto \xi^{1/4}$, 
as the log-log plot of Fig. \ref{epeak} indicates. 

The width of the $n=1$ spectrum does not seem to depend strongly on $\xi$ when $n=1$, but increases with increasing electric field $F$ for both $n=0,1$, as shown 
in Figure \ref{creationraten1} (a)  which superimposes the $n=1$ creation rates for $F/m^2=0.1$ and $F/m^2=0.4$.

Fig. \ref{upsilonnumeric} shows log-log plots of the conversion coefficient $\Upsilon$ versus $\xi$. 
Notice the linear behavior of Log$_{10} \Upsilon$ in Figure \ref{upsilonnumeric}, which illustrates power-law dependence on $\xi$. Notably, the analytical estimate in Eq. \eqref{Ups-quant-g=0} confirms this behavior. 
However, the nonlinear behavior of the numerical data appearing at smaller values of $\xi$ is not reproduced by the analytical formula.  The reason is that the estimate was arrived at 
by integrating over all values of $p$---at smaller $\xi$, there exist fewer values of $p$
for which the $n=1$ collision occurs.  The $n=0$ collision occurs for all values of $p$, and so does not drop off in the same way.

A somewhat unexpected feature of these formulae
is observed in the plots of $\Upsilon$ for $n=0$  shown in Fig. 6 (a), at smaller values of $\xi$. 
For $\xi \leq 10^{3}$, the conversion coefficient exhibits a plateau that disappears at strong fields. 
For $F/m^2\sim 0.1$, the conversion coefficient changes minimally from $\xi=10$ to $\xi=10^{4}$. This is due to the absence of the large logarithm in the Hessian when $n=0$, see Eq. \eqref{D}. Thus, the energies of the $n=0$ recollision may be increased without a corresponding drop in the probability of recollision.

Taken all together, these calculations show the photon spectrum to possess a complex structure, with broad spectrum of photon energies $\Omega$ emitted
in the region $m<\Omega < m\xi$, but also a set of narrower peaks resulting from the ``caustic--type'' collisions.


\newpage
\newpage
{\color{white}followed by a red fragment}
\newpage
\section{Muon production}
\label{Muon production}
Calculating the conversion coefficient $\Upsilon$ for muon production is a more sophisticated problem. For charged particles such as the muon, the particle's energy in the final state depends on the time dependent electric field,  unlike in the photon case where $\Omega$ is time independent.

In place of \eqref{dWdrho}, charged particles are described by 
\begin{align}
\sigma(\varepsilon_\mu)=\frac{2\pi}{v_{12}}\int {\sum}^\prime |M(\varepsilon_\mu)|^2~\frac{d\rho}{d\varepsilon}~d\mathcal{O} ,
\label{musigma}
\end{align}
where $\mathcal{O}$ denotes the solid angle of created muons and $\varepsilon_\mu$ is the muon energy. Hence, 
\begin{align}
\frac{dW}{d\varepsilon d\mathcal{O}}=
\frac{{F}^{1/2}\, }{4\sqrt{2}\,\xi}\frac{d\sigma(\varepsilon)}{d\mathcal{O}}
\sum_{n\ge 0}\,
\frac{\,|Q|^2}
{|D(p)|^2}\,\exp\big(\!-\frac {\pi p^2}{2\,\xi^2F}\,\big),
\label{mudWdeps}
\end{align}
For photons, integration of Eq. \eqref{mudWdeps} over $\mathcal O$ was trivial. In the muon case the factor of $|Q|^2 $ depends on the angle between the muon momentum and the electric field, as it depends on $\left(\boldsymbol{p}_\mu+ \tfrac{\boldsymbol{F}}{\omega}\sin\omega t\right)^2$, so that integration over $\mathcal O$ to acquire the spectrum, and $\varepsilon$ to acquire $\Upsilon$, becomes difficult. We leave the solution of this problem to future analytical and numerical studies.

\begin{figure}[htp]
\flushleft
\vspace{0.2cm}{
\includegraphics[scale=0.56]{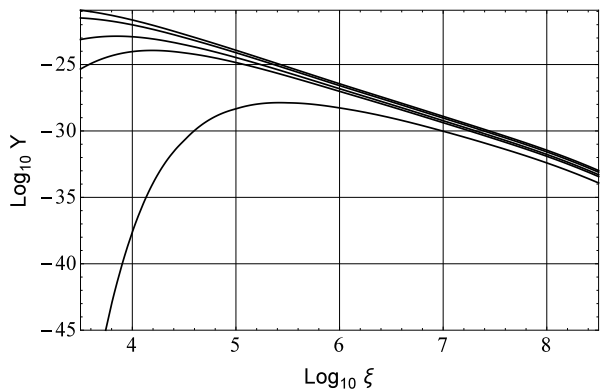}
}
\caption{Conversion coefficient for uncharged muons. The curves correspond to n=0, $F/m^2=0.01, \, 0.1, \, 0.2, \,0.5, \, 0.8$ (bottom to top).}
 \label{muonestimate}
\end{figure}

However, as a first approximation, we calculated $\Upsilon$ for ``uncharged muons'', calculating $\Upsilon$ through the same procedure used in the previous section on $2\gamma$ production, except with the appropriate cross section and the requirement that the collision energy be at least the muon mass, $\varepsilon(\tau)\geq m_\mu=105.7$ MeV. The result is presented in Figure \ref{muonestimate}. This probability is of the same magnitude as the probability of muon production found by Meuren et. al. \cite{Meuren2015}, but this work did not account for the suppression due to the spin polarization of the $e^-e^+$ pair, so in fact the recollision rate we consider, in a laser standing wave, is likely to be several orders of magnitude larger than for the collision of a photon and plane wave considered in \cite{Meuren2015}.

\newpage
\newpage
\section{Discussion}
\label{discussion}
The creation rates and spectra of high energy photons and heavy particles produced by two 
colliding laser beams are discussed. When the field is sufficiently strong, $e^-e^+$ pairs are created through the
Schwinger mechanism. Our observation is that the same field can accelerate the fermions 
and place them on a collision course.
The resulting $e^-e^+$ collision produces high energy particles in the final state of the reaction. 
Specifically, the creation of gamma quanta,  $e^+e^- \rightarrow 2\gamma$, and muon production,
$e^+e^-\rightarrow \mu^+\mu^-$, were discussed.

The process considered is based on the concept of the antenna mechanism, which describes how a variety of different systems may efficiently absorb energy from a strong, low-frequency laser field \cite{Kuchiev1987,Kuchiev2007}.
However, there is an interesting and important distinction with the previously considered antenna--type applications.
The colliding laser beams produce an almost homogeneous electric field in the laser focus.
As a result, the $e^-e^+$ pair creation and propagation takes place as a coherent quantum process, at all points within the laser focus at $\omega t=\omega x=0$.

This is different from the problems considered previously, when an electron is created in the vicinity of an atom \cite{Kuchiev1987,Lewenstein1994,Corkum1993}, or when an $e^-e^+$ pair is created near a nucleus \cite{Kuchiev2007}, the fermions are created in the near vicinity of the atom/nucleus. In those cases, the created particles possess well-defined coordinates, whereas in the present paper they possess well--defined momenta, leading to enhanced yields. Those previous studies rely on semiclassical treatments in which the classical trajectories emanate from the coordinates of the atomic/nuclear particle, by contrast in the problem at hand one cannot indicate any special coordinate responsible for the $e^-e^+$  creation.

Hence from the very beginning the process required the coherent quantum description, making the problem more theoretically involving. We needed to construct and use the time--dependent wave function for the $e^-e^+$ pair, which describes the amplitude of pair creation and its subsequent propagation within the whole volume exposed to the strong laser field.

The coherent nature of the problem has a remarkable consequence. We found that it produces a strong boost for the probability of the recollision, and therefore provides a substantial additional enhancement of high energy particle production. This fact is demonstrated through the analytical and numerical calculations.

\begin{figure}[b]
\flushleft
\vspace{0.2cm}
\hspace*{-0.3cm}{
\includegraphics[scale=0.68]{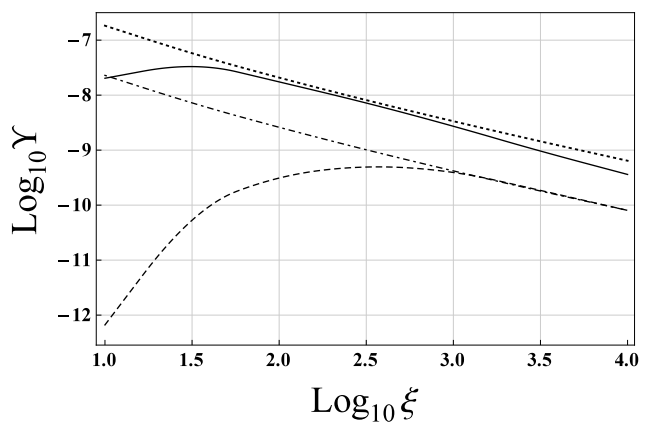}
}
\caption{The conversion coefficient $\Upsilon$, i.e. the number of photons produced per $e^-e^+$ pair, versus the adiabatic parameter $\xi$.
The analytical result Eq. \eqref{Ups-quant-g=0-n=1} and numerical calculations --- 
the dotted (analytical) and solid (numerical) lines: $F/m^2=0.8$; 
dot--dashed (analytical) and dashed (numerical) lines: $F/m^2=0.1$. }
 \label{estimation}
\end{figure}

The time dependent, many-body, quantum nature of the problem makes it rather involving.
It is therefore surprising that it was possible to derive certain analytical statements.
Comparison between our analytic and numerical results
is illustrated in Fig. \ref{estimation}, which presents 
the conversion coefficient for the gamma production (number of photon pair per one $e^+e^-$ pair)
versus the adiabatic parameter $\xi=e\mathcal E/m\omega$. The figure shows good quantitative and semi--qualitative agreement between the numerical and analytical results.

As is typical of the WKB--type semiclassical approach, the classical trajectories of the $e^-$ appear. We find that the necessary classical trajectories progress linearly along the direction of the electric field. This is unsurprising, as it is well--known that in quantum electrodynamics that the vacuum exhibits to the dichroism of the vacuum in quantum electrodynamics; the vacuum predominantly absorbs light polarized along the direction of the electric field, which means that the $e^-e^+$ pairs produced are similarly polarized \cite{Klein1964}.

The numerical calculations show that the spectrum of emitted gamma quanta
possesses two distinctive features. Firstly note that 
the larger the adiabatic parameter $\xi = F/m\omega$, the larger is the energy available to
 each fermion in the laser field,  $\varepsilon(t)\propto m\xi$. 
At the moment of collision, this large energy can be transformed into the energy of the emitted photons, so the spectrum includes higher photon energies as $\xi$ is increased. 
The lower limit on the photon energy is given by the electron mass. 
One ought to anticipate therefore an existence of a broad spectrum of photon energies $\Omega$ emitted
in the region, $m<\Omega < m\xi$.
Our calculations support these expectations.

Secondly, alongside the smooth energy distributions of emitted photons
there also exist much narrower, pronounced maxima in the spectrum resulting from the ``caustic--type'' collisions.
The electron (positron) energy $\varepsilon(t)$ in the laser field is a function of time; we found that the probability of the $e^+e^-$ collision is boosted
if at the moment of collision $t=\tau$ when the energy derivative
turns zero, $\frac{d}{dt}\,\varepsilon(\tau)=0$. The intensity, position, width and sophisticated shape of the resulting maxima in the photon spectrum depend on the intensity and frequency of the field, enriching the structure of the photon spectrum.

The efficiency of the antenna mechanism must 
diminish with the increase of $\xi$, since in the static limit the electric field cannot
put the pair on a collision course.
Interestingly, our calculations show that this decrease in $\Upsilon$ is logarithmically slow,  
$\propto 1/\mathrm{ln}^2\xi$. 
This means that the antenna mechanism produces an intense flux of colliding $e^+e^-$ fermions
even at large $\xi$, when the photon energy is high. 
The origin of this efficiency can be traced down to the focussing of the fermion trajectories along the electric field.

On the other hand, the relevant cross section shows the standard power-type decrease
with increasing electron energy. For example, for the photon production 
$\sigma(\varepsilon)\approx \pi r_e^2m^2/\varepsilon^2\propto \xi^{-2}$, 
which gives sizable suppression for high energy photon production. 
However, this feature is not unique to the
collisions considered  ---  it manifests itself similarly in any process where high energy 
$e^-e^+$ collisions are present.

Qualitatively similar observations can be made in relation to the production of muon pairs.
Unlike the photon case, the $\mu^+\mu^-$ pairs can be created directly,
via the Schwinger mechanism. However, the antenna-type production, via the intermediate $e^-e^+$ pair,
exponentially prevails over the Schwinger process; 
the ratio of the two probabilities is  $\propto \exp(\pi m_\mu^2/F)$, which is enormous. 

The mentioned exponential enhancement does not remove
all obstacles to the observing muon production in laser fields.
Despite the removal of the exponential suppression, the process remains suppressed by the preexponential factor present in the cross section of the $e^+e^-\rightarrow \mu^+\mu^-$ collision. 
Even for small muon energies in the final state, this cross section remains small $\sigma(\mu^+\mu^-)\propto r_e^2 
(m/m_\mu)^4$, producing sizable suppression for the muon yield. For unpolarized muons, the cross section would contain the factor $(m/m_\mu)^2$ --- it is the opposite spirality states of the $e^-$ and $e^+$ which introduce the additional suppression \cite{Berestetskii1982}.

To cite some numerical values for the processes considered, take the case
of gamma production
and tune the field to the maximum value $F=m^2$, which is on the border of the region allowed by the 
approximations considered. Figure \ref{estimation} shows that the coefficient of conversion reaches the value $\Upsilon\approx 10^{-7}$
for $\xi\approx 10^{1.5}$. 
This is an encouraging result, as it greatly exceeds the rate of photon production in the collision of a laser beam and a nucleus \cite{Kuchiev2007}.

The calculated values of $\Upsilon$ in Fig. \ref{upsilonnumeric} are encouraging. An electric field corresponding to a peak intensity of $10^{27}$ Wcm$^{-2}$, yields a Schwinger-type exponential suppression of the $e^-e^+$ creation rate of order $\mathcal O(10^{-20})$, but can be compensated for by the comparably large four-volume of the focal spot of the laser $\left(\lambda_{laser} / \lambda_c \right)^4 \approx 10^{22}$, where $\lambda_c$ is the Compton wavelength i.e. even for small creation rates, pair creation is observable \cite{ELI2011}. Larger laser focii, and other techniques such as the dynamical Schwinger mechanism, may similarly compensate for the apparently small value of $\Upsilon$. Moreover, the volume over which photons may be detected can be made much larger than the laser focus, since the photons may propagate outward away from the point of recollision, across distances much greater than the laser frequency.

Production of photons with energies $\geq $ 100 GeV seems plausible with the lasers of the future, with $F/m^2=0.3$ and $\xi \sim 10^{5}$. More immediately, observation of pair production can be made feasible at ELI with $F/m^2=0.1$ and $\xi=5\cdot 10^4$ (as suggested by their documentation \cite{ELI2011}); the conversion coefficient can be of the order of $\Upsilon \sim 10^{-10}$, and the energies of the photons produced may reach $m\xi\sim$ 25 GeV \cite{Dunne2009}.

However, $\Upsilon=10^{-10}$ nonetheless presents a challenge for experimental observation. 
Searching for possible enhancement, it is tempting to contemplate development of the approach presented 
here into the region of strong fields, $F > m^2$. We do not make such an attempt here, but the appeal of such extension should be kept in mind
for future studies. In such a region, the radiation reaction would most likely become significant, as $\chi^2 > 1$, making avalanches of $e^-e^+$ and $\gamma$ cascades possible. The problem then resembles high--harmonic generation in a plasma, the plasma consisting of a swarm of pairs created by the strong field.

The possible observations of the effects considered need certain experimental conditions to be satisfied.
First of all, the laser intensity should be large while the laser frequency remains low to make the 
$e^+e^-$ production probable and supply high wiggling energy for these fermions.
Hopefully, in the next generation of laser facilities these conditions would be satisfied. As mentioned previously, the parameter $\xi$ 
can be made very large in experimental facilities scheduled 
to be operational in the next few years, 
making our process 
capable of producing photons with energies of up to 25 GeV
and higher in the immediate future.
Secondly, the two colliding laser beams need frequencies, polarizations and
phases to be tuned to produce a standing wave. 
It may be a challenging task, but the outcome looks very attractive.
With the projected intensities and frequencies of laser beams, laser facilities
of the next generation would plausibly be capable of studying $e^+e^-$ collisions 
at energies close to, even exceeding 100 GeV.

% \begin{figure}[b]
%\flushleft
%{\includegraphics[scale=0.30]{phiE2.pdf}}
%\caption{Photon spectrum for $n=0$ 
%as a function of the electron energy $\varepsilon$ for $\xi=5\cdot 10^{4}$;
%bottom to top: $F/m^2$ =  $ 0.0125, \ 0.025, \ 0.05, \ 0.1, \ 0.2, \ 0.4, \ 0.8$.}
 %\label{energy}
%\end{figure}

\newpage
\appendix
\section{Adiabatic approximation}
\label{Adiabatic approximation}
We derive an approximation to the electron wave function, valid when the wave function varies slowly with respect to the field.
Consider the electron-positron pair creation by a periodic vector potential $A(t)$ and corresponding
electric field $\mathcal E$,
which are aligned along the $z$ axis and satisfy
\begin{align}
e A(t)=\frac{F}\omega \sin \omega t~,
\label{Aot}
\\
e \mathcal E(t)=F \cos \omega t~.
\label{Fomegat}
\end{align}
Let $\psi^{ (\pm) }_{\mathbf{p}}(\mathbf r,t)$ be the electron and positron wave functions---the subscript indicates the sign of the particle's charge (i.e. $-$ for the electron and $+$ for the positron). The Dirac equation reads
\begin{align}
i\frac{\partial \psi^{ (\pm) }_{\mathbf{p}}}{\partial t}= \Big(\mp\boldsymbol{\alpha} \cdot(\boldsymbol{p}-e\boldsymbol{A}(t)) + \beta m \Big)\psi^{ (\pm) }_{\mathbf{p}}~.
\end{align}
Introducing the adiabatic wave function 
\begin{align}
\psi^{ (\pm) }_{\mathbf{p}}(\mathbf r,t)=
 \varphi^{ (\pm) }_{\mathbf{p}}(\boldsymbol{r}, t) \
\exp\Big(i \int^t\varepsilon(t') \ dt'\,\Big)~,
\end{align}
substitution into the Dirac equation gives
\begin{align}
(\boldsymbol{\alpha} \cdot(\boldsymbol{p}\pm e\boldsymbol{A}(t)) + \beta m)~\varphi^{ (\pm) }_{\mathbf{p}}(\boldsymbol{r}, t)=\varepsilon(t)~ \varphi^{ (\pm) }_{\mathbf{p}}(\boldsymbol{r}, t)
\end{align}
and so
\begin{align}
\varphi_\pm = \chi^{ (\pm) }_{\mathbf{p}(t)} \ \exp\Big(\pm i\left(\boldsymbol{p} \pm e\boldsymbol{A}(t)\right)\cdot \boldsymbol{r}\,\Big)~,
\end{align}
is a solution if we neglect the partial derivatives of the spinor $\chi^{ (\pm) }_{\mathbf{p}(t)}$ 
\begin{align}
\frac{\partial \chi^{ (\pm) }_{\mathbf{p}(t)}}{\partial\boldsymbol{r}}=\frac{\partial \chi^{ (\pm) }_{\mathbf{p}(t)}}{\partial t}=0.
\end{align}
This is the content of the adiabatic approximation. We have solved the Dirac equation as if the time-dependent field is at some fixed moment in time $t=t'$.
The spinors $\chi^{ (\pm) }_{\mathbf{p}(t)}$ are given by
\begin{equation}
(\boldsymbol{\alpha}\cdot \boldsymbol p (t)+\beta\, m)~\chi^{ (\pm) }_{\mathbf{p}(t)}\,=\,\varepsilon(t)~ \chi^{ (\pm) }_{\mathbf{p}(t)}~.
\label{u-}
\end{equation}
At any given moment of time $t$ the spinor $\chi^{ (\pm) }_{\mathbf{p}(t)}$ can thus be considered as describing a plane wave with
momentum $\boldsymbol p (t)=\boldsymbol{p} -e\boldsymbol{A}(t') $ and energy $\varepsilon(t)$, as noted in Section IV A. These wave functions incorporate the laser-fermion interaction to all orders.

We choose the usual normalization of of one particle per volume,
\begin{equation}
\overline{\chi}_{\mathbf{p}(t)}^{(\pm)}\chi^{ (\pm) }_{\mathbf{p}(t)} =2m.
\label{uu}
\end{equation}
 \newpage

\section{Keldysh method}
\label{Keldysh method}
Consider the electron-positron pair creation by a periodic vector potential $A(t)$ and corresponding
electric field $\mathcal E$,
which are aligned along the $z$ axis and satisfy \eqref{Aot} and \eqref{Fomegat}. The amplitude of the pair creation is given by
\begin{align}
\mathcal{A} &= \frac{\omega}{2\pi}\int^{\frac{2\pi}{\omega}}_0 \bra{\psi^{ + }_{-\mathbf{p}}}-e\boldsymbol{A}\cdot\boldsymbol{\alpha}\ket{\psi^{ - }_{\mathbf{p}}} \ dt \nonumber \\
&= -\frac{\omega}{2\pi}\int^{\frac{2\pi}{\omega}}_0 \bra{\varphi_+}e\boldsymbol{A}\cdot\boldsymbol{\alpha}\ket{\varphi_-} \ e^{-2i\int^t \varepsilon (t') \ dt'} \ dt
\label{amplitude}
\end{align}
see e.g. Ref. \cite{Kuchiev2007a}. Representing the amplitude $\mathcal{A}$ of the pair creation with exponential accuracy,
\begin{equation}
\mathcal{A} \propto e^{iS},
\label{PsiiS}
\end{equation}
the relevant part of the semiclassical action $S$ may be evaluated using the Keldysh method \cite{Keldysh1965}
\begin{align}
S=-2\int_C \varepsilon (t)dt~.
\label{SKeldysh1965}
\end{align}
Here $\varepsilon(t)$ is the electron energy
\begin{equation}
\varepsilon(t)=\pm \Big(m^2_e+p_\perp^2+\big(p_\parallel-e A(t)^2\big)\Big)^{1/2},
\label{energy(t)}
\end{equation}
where $p_\perp,p_\parallel$ are the 
components of the electron momenta
at the moment of its creation ($t=n\pi/\omega$, where
the field reaches its maximum) perpendicular and parallel to the field.
To be specific, we assume this point to be  $t=0$.
The contour $C$ in (\ref{SKeldysh1965}) runs in the upper semiplane of 
the complex plane of time $t$ starting and finishing at 
the point of creation $t=0$.
At the initial point on this contour the energy 
$\varepsilon(t)$ belongs to the lower-continuum
described by the minus sign  in (\ref{energy(t)});
at the final point on the contour the energy belongs to the 
conventional upper-continuum specified by the plus sign in (\ref{energy(t)}).
To make the transition from one continuum to another the 
contour $C$ runs clockwise the upper half-plane of $t$, surrounding the branch point $t_0$ of the energy, which is located where the condition
\begin{equation}
\varepsilon (t_0)=0
\label{energy=0}
\end{equation}
is satisfied.
According to these definitions, the action can be rewritten
\begin{equation}
S=2\int_0^{t_0} \Big(m^2_e+p_\perp^2+\big(p_\parallel+(F/\omega)\sin \omega t \big)^{2}\Big)^{1/2}dt~.
\label{Srewrite}
\end{equation}
By scaling the parameters $p_\perp,p_\parallel$ 
and introducing a new integration variable $x$,
\begin{align}
&Q=p_\perp/m,\quad\quad P=p_\parallel/m,
\\
&x=\xi\sin \omega t,
\end{align}
we simplify the expression for the action
\begin{equation}
S=\frac{2m}F \int_0^{X}\left(\frac{1+Q^2+(P-x)^2}{1-x^2/\xi^2}\right)^{1/2}dx~.
\label{elliptic}
\end{equation}
Here 
\begin{equation}
X=P+i(1+Q^2)^{1/2}~.
\label{XQP}
\end{equation}
The integral in (\ref{elliptic}) is of a conventional elliptic form.
In order to simplify it we can
rely on the following estimates for the relevant momenta 
\begin{equation}
|Q| \lesssim \sqrt F/ m,\quad\quad |P|\lesssim \,{\xi\sqrt F}/  m,
\label{Q,P<}
\end{equation}
which are justified at the end of this appendix.
According to (\ref{weakfield}) and (\ref{xi}) these inequalities imply that $|Q|\ll 1$, $|P|/\xi\ll 1$.
We derive from this that on the whole contour $C$ the integration variable $x$ in (\ref{elliptic}) satisfies $ |x|^2/\xi^{2}\lesssim F/m^2\ll 1$, which allows us to expand the denominator in the integrand in (\ref{elliptic})
\begin{equation}
1/(1- x^2/\xi^{2})^{1/2}\approx 1+ x^2/(2\xi^{2}).
\label{approx1/2}
\end{equation}
After this expansion, the integral in (\ref{elliptic}) is evaluated in terms of elementary functions
\begin{align}
&S=\frac{m^2}F
\bigg\{
\frac{i\pi}{2}
(1+Q^2)\Big(1-
\frac{1}{8\xi^2}(1-4P^2
+Q^2)\Big)-
\\
&\Big[
P(1+P^2+Q^2)^{1/2}
\Big(1-\frac{1}{24\xi^2}(13-2P^2+13Q^2)\Big)+
\nonumber
\\
&(1\!+\!Q^2)\Big(1\!-\!\frac{1}{8\xi^2}(\!1\!-\!4P^2\!+\!Q^2)\!\Big)
\!\ln\frac{P\!+\!(1\!+\!P^2\!+\!Q^2)^{1/2}}{(1+Q^2)^{1/2}}
\Big]
\!\bigg\}.
\nonumber
\end{align}
Limiting these functions
to the region of interest (\ref{Q,P<}), we arrive
at a transparent result for the action $S$,
which yields the amplitude of the pair creation \eqref{PsiiS}
\begin{align}
%&S =i A+B~,
%\label{iSiAB}
%\\
&\mathrm{Im}\, S=\frac{\pi m^2}{2F}\Big(1+Q^2+\frac{1}{2\xi^2} P^2\Big)~,
\label{AfromS}
\\
&\mathrm{Re}\,S=-\frac{m^2}F
\times
\left\{
\begin{array}{lc}
\int_0^{P}\left(1+P'^2\right)^{1/2} \ dP', & 1\ll |P|\ll \xi
\\
2P,  & |P|\ll 1
\end{array}
\right. 
\label{BfromS}
\end{align}
%(A simlpe interpolating formula for $B$ reads $B\approx -(m^2/F) P(1+\sqrt{1+P^2+Q^2}$.)
According to (\ref{PsiiS}), the real part of the action gives the phase factor of the amplitude, while
its imaginary part defines the probability of the pair creation with  exponential accuracy,
\begin{equation}
%= |\Psi|^2\propto \left|e^{2iS}\right|
w \propto e^{-2\mathrm{Im}\,S}
=\exp\Big[-{\frac{\pi m^2}{F}\Big(1+Q^2+\frac1{2\xi^2} P^2\Big)\Big]}.
\label{w}
\end{equation}
This expression shows that the probability remains sizable only within the region restricted by inequalities (\ref{Q,P<}),
which justifies our interest in this specific region.
 \\
 \\
 
\section{Third order contribution to action}
\label{third order}
Expanding the function $\tilde S$ around the saddle point, at which $\partial_t \tilde S = 0$,
\begin{align}
\tilde S &= 2\Omega + S(p,0,\tau) +\frac{1}{2}\tilde S ''(p,0,\tau) (t-\tau)^2 \nonumber \\
& \ \ \ \ \ \ \ \ \ \ \ \ \ \ \ \ \ +\frac{1}{6}\tilde S '''(p,0,\tau) (t-\tau)^3+... \nonumber
\end{align}
Simple calculations yield
\begin{align}
\tilde S'''(p,0,\tau) &= -2\frac{F^2}{m}\left(\frac{m^3}{\varepsilon^3(\tau)}w\left(2w-\cos\omega \tau \right) \right. \nonumber\\
& \ \ \ \ \ \ \ \ \ \ \ \ \ \ \left. -\frac{m^3}{\varepsilon^3(0)}\frac{v^3}{(2n+v(0))^3}-\xi^{-1} v \sin \omega \tau \right).
\end{align}
Using the integral for the Airy function Ai(x), up to a phase we find
\begin{align}
& \int_{-\infty}^\infty \exp \left(-i\tilde S \right) \ dt  \nonumber \\
&= i\frac{2^{4/3}\pi}{\tilde S '''(p,0,\tau)^{1/3}} \text{Ai} \left(-\frac{\tilde S ''(p,0,\tau)^2}{2^{2/3}\tilde S '''(p,0,\tau)^{4/3}}\right),
\end{align}
and, as $\varepsilon(\tau)\sim m\xi $, conclude (ignoring $n=0$ for simplicity)
\begin{align}
\frac{dW}{d\varepsilon}& = \sum_{n\ge 1} \ \frac{\pi^2\alpha^2\sqrt{F}}{8\sqrt{2}\left(\log\pi\xi \right)^2}\frac{\exp\big(-\frac {\pi p^2}{4\,\xi^{2}F}\big)}{n^2\,(2n+v(0))}~\sigma(\varepsilon)\nonumber \\
& \ \ \ \ \ \ \ \ \ \times X(\tau)\Big(\text{Ai}\left(-X(\tau)~w(\tau)^2 \right) \Big)^2~,
\end{align}
where
\begin{align}
X(\tau)=\Big(\frac{\xi^{2}|v(\tau)|}{ F |\sin\omega\tau |}\Big)^{2/3}~.
\end{align}
\section{Van Vleck Determinant}
\label{VVD}
Calculation of the $e^-e^+$ wave function may be equivalently be thought of in terms of the van Vleck propagator \cite{VanVleck1928}. This semiclassical expression for the electron wave function is 
\begin{align}
\Psi(\boldsymbol{R},t)=\frac {1 }{(2\pi i)^{3/2}}\text{det}\left(\frac{\partial^2\bar{S}}{\partial R_i\partial  r_k}\right)^{1/2} e^{iS_0}\Psi(\boldsymbol{r},0)
\label{vvmethod}
\end{align}
where the right hand side contains the wave function at zero $t=0$, and at the initial coordinate $\boldsymbol{r}$, the action $S_0$ on the classical trajectory from $\boldsymbol{r}$ to $\boldsymbol{R}$ When the pair recollide, we consider in the case where $\boldsymbol{R}=\boldsymbol{r}$.

Our calculation of the wave function essentially reads
\begin{align}
\Psi(0,t)&=\frac {1 }{(2\pi i)^{3/2}}  \text{det}\left(\frac{1}{2}\frac{\partial^2 S}{\partial p_i\partial p_k}\right)^{-1/2}\,e^{i S_{0}} \Psi(\boldsymbol{r},0)
\label{saddlemethod}
\end{align}
It is easy to see these two expressions \eqref{vvmethod} and \eqref{saddlemethod} are equivalent provided
\begin{align}
\text{det}\left(\frac{\partial^2\bar{S}}{\partial R_i\partial  r_k}\right) = \text{det}\left(\frac{1}{2}\frac{\partial^2 S}{\partial p_i\partial p_k}\right)^{-1}
\label{lemma}
\end{align} 
We shall prove this lemma below. It is useful to make clear from the outset that there is a distinction between the action in Eq. \eqref{vvmethod} and Eq. \eqref{saddlemethod}; while $\bar{S}=\bar{S}(\boldsymbol{R},\boldsymbol{r},t)$, on the other hand $S=S(\boldsymbol{p},\boldsymbol{r},t)$. Deviating somewhat from the notation in the text, we denote the electron energy $\varepsilon=\varepsilon(\boldsymbol{p})$; then explicitly we have
\begin{align}
\bar{S}(\boldsymbol{R}, \boldsymbol{r}, t) &= S(\boldsymbol{p}, \boldsymbol{r}, t) -\boldsymbol{p} \cdot \boldsymbol{R} \nonumber \\
&= \boldsymbol{p}\cdot(\boldsymbol{r}-\boldsymbol{R}) -\int_0^t \varepsilon\left(\boldsymbol{p}+\boldsymbol{F}/\omega\sin\omega t'\right) \ dt' \nonumber
\end{align} 
The saddle point condition reads
\begin{align}
\frac{\partial \bar{S}}{\partial \boldsymbol{p}}=\frac{\partial S}{\partial \boldsymbol{r}}-\boldsymbol{R}=0, \nonumber
\end{align}
 implying the classical trajectory
\begin{align}
\boldsymbol{r}=\boldsymbol{R}+\int_0^t \boldsymbol{v}\left(\boldsymbol{p}+\boldsymbol{F}/\omega\sin\omega t'\right) \ dt'. \nonumber
\end{align}
Restricting ourselves to motion along this trajectory implicitly defines $\boldsymbol{p}=\boldsymbol{p}(\boldsymbol{R},\boldsymbol{r},t)$. Simple applications of the chain rule show that
\begin{align}
\frac{\partial^2 S}{\partial R_i \partial r_i}&= -\frac{\partial p_i}{\partial r_k}+\frac{\partial p_j}{\partial R_i}\frac{\partial }{\partial r_k}\left(\frac{\partial S}{\partial p_j}-R_i \right) \nonumber\\
&=-\frac{\partial p_i}{\partial r_k}+\frac{\partial p_k}{\partial R_i}
\label{key}
\end{align}
Differentiating the expression for the classical trajectory, it follows that 
\begin{align}
\delta_{ij} &=\frac{\partial p_k(\boldsymbol{r},\boldsymbol{R},t)}{\partial r_j} \int_0^t \frac{\partial\boldsymbol{v}}{\partial p_k(\boldsymbol{r},\boldsymbol{R},t)} \  dt' \nonumber \\
&= -\frac{\partial p_k}{\partial r_j}\frac{\partial^2 S}{\partial p_i \partial p_k} \nonumber
\end{align}
We know from our earlier discussion that there exists a basis in which the Hessian is diagonal (when the electric field is aligned along the $z$-axis), so
\begin{align}
\frac{\partial^2 S}{\partial p_i \partial p_k }=\delta_{ij}\Sigma_i \nonumber
\end{align}
from which we find
\begin{align}
\frac{\partial p_k}{\partial r_j}=-\frac{\partial p_k}{\partial R_j}=-\delta_{jk} \Sigma^{-1}_j\nonumber
\end{align}
Using Eq. \eqref{key}, this last line becomes
\begin{align}
\frac{\partial^2 S }{\partial R_i \partial r_k}=-2\frac{\partial p_i}{\partial r_k}=2\delta_{ij} \Sigma^{-1}_i \nonumber
\end{align}
from which Lemma \eqref{lemma} immediately follows.

Underpinning the formula for the $e^-e^+$ wave function \eqref{explicit-wf} are a great deal of technical details---the trajectories, the integration over momenta, the classical action---they may all be conceptualised through the simple interpretation as a calculation of the van Vleck propagator. It so happens that the propagator is non--zero for $\boldsymbol{r}=\boldsymbol{R}$ and $t\neq0$, meaning the fermions to collide.

\section*{Acknowledgments} 
This work was supported by the Australian Research Council.

%\bibliography{ENER}{}

\begin{thebibliography}{66}

\bibitem{Sauter1931}
F.~ Sauter, { Zeitschrift f\"ur Physik} {\bf 82}, 742 (1931).

\bibitem{Schwinger1951}
J. Schwinger,
\newblock { Phys. Rev.}, \textbf{82}, 664 (1951).

\bibitem{Brezin1970}
E. Brezin and C. Itzykson, 
\newblock { Phys. Rev.} D, \textbf{2}, 1191 (1970).

\bibitem{Narozhnyi1973}
N. B. Narozhnyi and A. I. Nikishov, 
\newblock {Zh. Eksp. Teor. Fiz.} \textbf{65}, 862 (1973) [Sov. Phys. JETP \textbf{38}, 427 (1974)].

\bibitem{Marinov1977}
M. S. Marinov and V. S. Popov,
\newblock { Fortschr. Phys.}, \textbf{25}, 373, (1977).

\bibitem{Popov1971} V. S. Popov, Pis’ma Zh. Eksp. Teor. Fiz. \textbf{13} 261 (1971) [JETP Lett. \textbf{13} 185 (1971)]; Zh. Eksp. Teor. Fiz. \textbf{61} 1334 (1971) [Sov. Phys. JETP \textbf{34} 709 (1972)]; Zh. Eksp. Teor. Fiz. 62, 1248 (1972) [Sov. Phys.-JETP 35, 659 (1972)].

\bibitem{Yakovlev1966}
V.~ P.~ Yakovlev,
\newblock Zh. Eksp. Teor. Fiz. \textbf{49}, 318 (1965) [Sov. Phys. JETP \textbf{22}, 223 (1966)].

\bibitem{Muller2003} 
 C. M\"uller, A.B. Voitkiv and N. Grun, Phys. Rev. A \textbf{67}, 063407 (2003);
Phys. Rev. A \textbf{67}, 063407 (2003).

\bibitem{Kuchiev2007a}
M. Yu. Kuchiev and D. J. Robinson,
\newblock { Phys. Rev. A}, \textbf{76}, 012107 (2007).

\bibitem{Ringwald2001} A. Ringwald, Phys. Lett. B \textbf{510}, 107 (2001).

\bibitem{Muller2004} C. M\"uller, A. B. Voitkiv and N. Gr\"un, Phys. Rev. A \textbf{70}, 023412 (2004).

\bibitem{Shearer1973} J. W. Shearer, J. Garrison, J. Wong and J. E. Swain, Phys. Rev. A \textbf{8}, 1582 (1973).

\bibitem{Liang1998} E. P. Liang, S. C. Wilks and M. Tabak, Phys. Rev. Lett. \textbf{81}, 4887 (1998).

\bibitem{Muller2008}
C. M\"uller, K. Z. Hatsagortsyan and C. H. Keitel,
\newblock { Phys. Rev. A}, \textbf{78}, 033408 (2008).

\bibitem{Nerush2011}
E. N. Nerush, I. Yu. Kostyukov, A. M. Fedotov, N. B. Narozhny, N. V. Elkina and H. Ruhl,
\newblock { Phys. Rev. Lett.}, \textbf{106}, 035001 (2011).

\bibitem{Elkina2011}
N. V. Elkina, A. M. Fedotov, I. Yu. Kostyukov, M. V. Legkov, N. B. Narozhny, E. N. Nerush and H. Ruhl,
\newblock { Phys. Rev. ST Accel. Beams}, \textbf{14}, 054401 (2011).

\bibitem{Alkofer2001}
R. Alkofer, M. B. Hecht, C. D. Roberts, S. M. Schmidt and D. V. Vinnik,
\newblock { Phys. Rev. Lett.}, \textbf{87}, 193902 (2001).

\bibitem{Muller2008a} C. M\"{u}ller, A. Di Piazza, A. Shahbaz, T. J. B\"{u}rvenich, J. Evers, K. Z. Hatsagortsyan and C. H. Keitel, Laser Phys. \textbf{18}, 175 (2008).

\bibitem{Fried2001} H. M. Fried, Y. Gabellini, B. H. J. McKellar and J. Avan, Phys. Rev. D \textbf{63}, 125001 (2001).

\bibitem{Bamber1999} C. Bamber et. al., Phys. Rev. D \textbf{60}, 092004 (1999).

\bibitem{Narozhnyi2015} 
N. B. Narozhnyi and A. M. Fedotov, Contemporary Physics \textbf{56}:3, 249 (2015).

\bibitem{ELI2011} M. M.  Al\' eonard et. al., ELI White Book (THOSS, Berlin, 2011), \url{http://www.eli‑beams.eu/}.

\bibitem{xfeleu} ``European XFEL---Enlightening Science'' \url{http://www.xfel.eu/}.

\bibitem{hiperlaser} ``HiPER---Laser energy for the future'' \url{http://www.hiper-laser.org}.

\bibitem{Kohlfurst2013} C. Kohlf\"{u}rst, Mario Mitter, Gregory von Winckel, Florian Hebenstreit and Reinhard Alkofer, Phys. Rev. D \textbf{88}, 045028 (2013).

\bibitem{Kirk2009} J. G. Kirk, A. R. Bell and I. Arka, Plasma Phys. Cont. Fusion \textbf{51}, 085008 (2009).

\bibitem{DiPiazza2004} A. Di Piazza, Phys. Rev. D \textbf{70}, 053013 (2004).

\bibitem{Bell2008} A. R. Bell and J. G. Kirk, Phys. Rev. Lett. \textbf{101}, 200403 (2008).

\bibitem{Bulanov2010} S. S. Bulanov, T. Zh. Esirkepov, A. G. R. Thomas, J. K. Koga, and S. V. Bulanov, Phys. Rev. Lett. \textbf{105}, 220407 (2010). 

\bibitem{Schutzhold2008} R. Sch\"{u}tzhold, H. Gies, and G. Dunne, Phys. Rev. Lett. \textbf{101}, 130404 (2008).

\bibitem{DiPiazza2009} A. Di Piazza, E. L\"{o}tstedt, A. I. Milstein and C. H. Keitel, Phys. Rev. Lett. \textbf{103}, 170403 (2009).

\bibitem{Kuchiev2007} M. Yu. Kuchiev, Phys. Rev. Lett. \textbf{99}, 130404 (2007).

\bibitem{Meuren2015} S. Meuren, K. Z. Hatsagortsyan, C. H. Keitel and A. Di Piazza, Phys. Rev. Lett. \textbf{114}, 143201 (2015). 

\bibitem{Meuren2015a} S. Meuren, K. Z. Hatsagortsyan, C. H. Keitel and A. Di Piazza, Phys. Rev. D \textbf{91}, 013009 (2015).

\bibitem{Kuchiev1987} M.~ Yu.~ Kuchiev, Pis'ma Zh. Eksp. Teor. Fiz. \textbf{45}, 404 (1987) [Sov. Phys. JETP Letters, \textbf{45}, 319 (1987)].

\bibitem{Lewenstein1994} M. Lewenstein, Ph. Balcou, M. Yu. Ivanov, A. L'Huillier and P. B. Corkum, Phys. Rev. A \textbf{49} (1994).

\bibitem{Corkum1993} P. B. Corkum, Phys. Rev. Lett. \textbf{71}, 1994 (1993).

\bibitem{Kuchiev2000} M. Yu. Kuchiev and V. N. Ostrovsky, J. Phys. B: At. Mol. Opt. Phys. \textbf{34}, 405-430 (2000); \textbf{32}, L189-L196 (1999).

\bibitem{Kuchiev1999} M. Yu. Kuchiev and V. N. Ostrovsky, Phys. Rev. A \textbf{60}, 3111 (1999); \textbf{61}, 033414 (2000).

\bibitem{Shwartz2014} S. Shwartz, M. Fuchs, J. B. Hastings, Y. Inubushi, T. Ishikawa, T. Katayama, D. A. Reis, T. Sato, K. Tono, M. Yabashi, S. Yudovich, and S. E. Harris, Phys. Rev. Lett. \textbf{112}, 163901 (2014).

\bibitem{Smirnova2009} O. Smirnova, Y. Mairesse, S. Patchkovskii, N. Dudovich, Da. Villeneuve, P. Corkum and M. Yu. Ivanov, Nature \textbf{460}, 972 (2009).

\bibitem{Berestetskii1982} V. B. Berestetskii, E. M. Lifshitz and L. P. Pitaevskii, {\em ``Quantum Electrodynamics (Landau and Lifshitz Course of Theoretical Physics Volume 4) Third Ed.''}, Elsevier Butterworth Heinemann, Oxford, (1982).

\bibitem{Dunne2009} G. V. Dunne, The European Physical Journal D \textbf{55}, 2, 327 (2009).

% review of lasers currently being developed esp this chart with values of xi. referenced in the discussion as well as in the assumptions

\bibitem{DiPiazza2012} A. Di Piazza, C. M\"{u}ller, K. Z. Hatsagortsyan and C. H. Keitel, Rev. Mod. Phys. \textbf{84}, 1177 (2012).

% this is a fantastic review 

\bibitem{Heinzl2011} T. Heinzl, {\em ``Strong-Field QED and High Power Lasers''}, Plenary talk QFEXT11 Benasque Conference, \href{http://arxiv.org/abs/1111.5192}{arXiv:1111.5192} [hep-ph] (2011).

\bibitem{Delone2000} N. B. Delone and V.P. Krainov, ``Multiphoton processes in atoms'', 2nd Ed. Springer-Verlag Berlin Heidelberg New York, (2000).	

% explanation of the averaging of the schwinger rate

\bibitem{Keldysh1965} L. V. Keldysh, Zh. `Eksp. Teor. Fiz. \textbf{47}, 1945 (1964) [Sov. Phys. JETP \textbf{20}, 1307 (1965)].

\bibitem{Popov1973} V. S. Popov, Zh. Eksp. Teor. Fiz. \textbf{63}, 1586 (1972) [Sov. Phys. JETP \textbf{36}, 840 (1973)].

\bibitem{Feynman2005}
 R.~P.~ Feynman, R.~B.~ Leighton and M.~Sands,
{\em ``The Feynman lectures on physics'', Volume 3}, Second Ed., Addison Wesley (2005).

\bibitem{Gribakin1997}
G. F. Gribakin and M. Yu. Kuchiev,
Phys. Rev. A \textbf{55}, 3760 (1997).

\bibitem{Copson2004}
E.~T.~Copson, {\em Asymptotic expansions}, Cambridge university press, 2004.

\bibitem{Seipt2015} D. Seipt, A. Surzhykov, S. Fritzsche and B. Kampfer, \href{arxiv.org/abs/1507.08868}{arXiv:1507.08868v1} [hep-ph] (2015).

% add to caustic discussion

\bibitem{Jackson1998} J. D. Jackson, {\em Classical Electrodynamics}, (Wiley, New York, 1998).

\bibitem{Nikishov1964} A. I. Nikishov and V. I. Ritus, Zh. Eksp. Teor. Fiz. \textbf{46}, 776 (1964) [Sov. Phys. JETP \textbf{19}, 529 (1964)].

\bibitem{Fedotov2010} A. M. Fedotov, N. B. Narozhny, G. Mourou and G. Korn, Phys. Rev. Lett. \textbf{105}, 080402 (2010).

\bibitem{Bell2008} A. R. Bell and J. G. Kirk, Phys. Rev. Lett. \textbf{101}, 200403 (2008).

\bibitem{DiPiazza2010} A. Di Piazza, K. Z. Hatsagortsyan and C. H. Keitel, Phys. Rev. Lett. \textbf{105}, 220403 (2010).

\bibitem{Landau1977} 
L.~D.~Landau and E.~ M.~ Lifshitz, {\em ``Quantum Mechanics: Non-Relativistic Theory, Volume 3, Third Ed."}, Elsevier Butterworth Heinemann, Oxford, (1977).

\bibitem{Page1957}
L.~ A.~ Page,
\newblock { Phys. Rev.}, \textbf{106}, 394 (1957).

\bibitem{VanVleck1928}
J.~H.~van Vleck, Proceedings of the National
Academy of Sciences USA {\bf 14}, 176 (1928). 

\bibitem{Klein1964}
J. J. Klein and B. P. Nigam,
\newblock { Phys. Rev.}, \textbf{136}, B1540 (1964).
 
 
\end{thebibliography}
%\bibliographystyle{nature}
%\bibliographystyle{h-physrev}
%\bibliographystyle{Nunsrt}
%\bibliographystyle{hunsrt}
%\bibliographystyle{h-physrev}
%plain
%nature
%is-unsrt
%unsrt
%Nunsrt
%nar
%phcpc
%phiaea
%hunsrt
%h-physrev
%\end{document}

\end{document}